\documentclass[12pt,preprint]{aastex}

\newcommand{\kepler}{{\it Kepler}}

\shorttitle{\kepler{} Mission Signal Detections}
\shortauthors{Tenenbaum et al.}

\begin{document}

\title{Detection of Potential Transit Signals in the First Three Quarters
of \kepler{} Mission Data}

\author{Peter Tenenbaum, Jessie L. Christiansen, Jon M. Jenkins, Jason F. Rowe, Shawn Seader,
Douglas A. Caldwell, Bruce D. Clarke, Jie Li, Elisa V. Quintana, 
Jeffrey C. Smith, Martin C. Stumpe, 
Susan E. Thompson, Joseph D. Twicken,
and Jeffrey Van Cleve}
\affil{SETI Institute/NASA Ames Research Center, Moffett Field, CA 94305, USA}
\email{peter.tenenbaum@nasa.gov}
\author{William J. Borucki, Miles T. Cote, Michael R. Haas, and Dwight T. Sanderfer}
\affil{NASA Ames Research Center, Moffett Field, CA 94305, USA}
\author{Forrest R. Girouard, Todd C. Klaus, Christopher K. Middour, and Bill Wohler}
\affil{Orbital Sciences Corporation/NASA Ames Research Center, Moffett Field, CA 94305, USA}
\author{Natalie M. Batalha}
\affil{San Jose State University, San Jose, CA 95192, USA}
\author{Thomas Barclay}
\affil{BAER Institute/NASA Ames Research Center, Moffett Field, CA 94305, USA}
\and
\author{James E. Nickerson}
\affil{Logyx Incorporated/NASA Ames Research Center, Moffett Field, CA 94305, USA}

\begin{abstract}
We present the results of a search for potential transit signals in
the first three quarters of photometry data acquired by the \kepler{} Mission.
The targets of the search include  
151,722 stars which were observed over the full interval and an additional
19,132 stars which were observed for only 1 or 2 quarters.  
From this set of targets we find
a total of 5,392 detections which meet the Kepler detection criteria: those criteria
are periodicity of the signal, an acceptable signal-to-noise ratio, and a composition
test which rejects spurious detections which contain non-physical combinations of
events.  The detected signals are dominated by events with relatively low signal-to-noise
ratio and by events with relatively short periods.  The distribution of estimated
transit depths appears to peak in the range between 40 and 100 parts per million, with
a few detections down to fewer than 10 parts per million.  The detections exhibit
signal-to-noise ratios from 7.1 $\sigma$, which is the lower cut-off for detections,
to over 10,000 $\sigma$, and periods ranging from 0.5 days, which is the lower cut-off
used in the procedure, to 109 days, which is the upper limit of achievable periods
given the length of the data set and the criteria used for detections.
The detected signals are compared to a set of known transit events in the
\kepler{} field of view which were derived by a different method using a longer data
interval; the comparison shows that the current search correctly identified 88.1\% of the
known events.  A tabulation of the detected transit signals, examples which illustrate
the analysis and detection process, a discussion of future plans and open, potentially
fruitful, areas of further research are included.
\end{abstract}



\section{Introduction}

The \kepler{} Mission \citep{wjb2010} uses a space-based photometer with a 115 $\deg^2$ 
field of view to search for transit signatures of extrasolar planets,
with a particular emphasis on Earth-size planets situated in the habitable
zones (HZs) of their parent stars.  Transit signatures are characterized by periodic
reductions in observed brightness as the planet blocks light
from the star\footnote{In the case of multi-body systems, for example multiple planets
orbiting a single star, gravitational interactions
can break the perfect periodicity of the orbits, leading to transit timing
variations (TTVs).  See Sections \ref{tps} and \ref{future} for more information
about the detection of planets which exhibit TTV.}.  
While Jovian-sized transiting exoplanets typically
produce a transit signature on the order of one percent in relative flux, 
which can be detected by relatively simple algorithms,
a true Earth-analog orbiting a Solar-type star is much more difficult
to detect:  its transits can result in brightness reductions of 100 parts per
million (PPM) or less, and the resulting reductions typically last for 1 to 13 hours
and repeat approximately once per year.  Given the large number of objects
which are observed by \kepler, the long datasets implied by the need 
for multi-year observation on sub-hour cadence, and the miniscule reductions in
intensity expected for small, rocky inner planets, one of the key data 
processing requirements for the \kepler{} Mission is an automated search for signals
which are consistent with transiting planets.  This is the task of the
Transiting Planet Search (TPS) software module \citep{jmj2002,jmj2010}.

This study describes the results of searching \kepler{}
observations acquired during the first
218 days of science operations with the TPS software module.

\subsection{\kepler{} Science Data}

The \kepler{} photometer maintains a constant pointing centered on 
$\alpha = 19^{\rm h}22^{\rm m}40^{\rm s}, \delta = +44.5\degr$.  In order
to permit nearly-continuous observation of its fixed field of view,
the \kepler{} photometer utilizes an Earth-trailing heliocentric orbit.
The requirement that \kepler{} maintain its solar panels pointing
at the Sun and its thermal radiator pointing away from the Sun implies that \kepler{} 
must rotate about its axis by $90\degr$ four times per solar orbit 
of 372 days.  The focal plane has been constructed with near-perfect $90\degr$
rotational symmetry to minimize disruption of the observations due to these rotations.
An observational period at a fixed axial rotation angle is referred to as a 
``quarter,'' and the maneuever from one rotation angle to another is referred
to as a ``quarterly roll.''

The data acquisition period selected for this analysis spans the first three
quarters (Q1-Q3) of science observation, spanning the interval from
2009 May 12 00:00:00 UTC to 2009 December 17 23:59:59 UTC, a total period of
218 days.  The first ``quarter'' consisted of approximately 33.4 days of
data acquisition following the end of spacecraft commissioning; the second and
third quarters each contain approximately 89 days of data acquisition.  Thus the
total acquisition time is 211.4 days, the balance being taken up by 
quarterly rolls and by a safe-mode event which was
immediately adjacent to a quarterly roll. 
During data acquisition, photometric data is taken with 6.0 second integrations, 
with each integration followed by a 0.5 second readout time;
these integrations are summed on board into 29.4 minute ``long cadence'' intervals.
The three quarters under discussion span a total of 10,363 such long cadences.  
In addition to the aforementioned data gaps in the
quarterly roll intervals, there are additional intra-quarter gaps due to 
spacecraft anomalies, reaction wheel desaturation (1 long cadence approximately
every 3 days), ``Argabrightening'' events\footnote{These are poorly-understood
events in which a large portion of the field of view will experience a modest 
increase in flux which lasts for 2-3 minutes; named for V. Argabright.} 
\citep{fcw2011}
and data downlink periods when the spacecraft must leave its science
orientation to point its high gain antenna at Earth (approximately 24 hours per
downlink, one downlink per month).  All told, these intra-quarter gaps consume
510 long cadences, leaving a total of 9,853 long cadence integrations containing
valid photometric data.

In order to limit the volume of downlinked data to acceptable levels, \kepler{} 
only acquires and downlinks photometric data for a small subset of pixels (i.e.,
5.44 million or fewer pixels out of 94.6 million pixels on the focal plane).  The pixels
are pre-selected in order to optimally capture the 
light from target stars which are expected to be of potential interest 
\citep{stb2010,nmb2010}.
A total of 170,854 target stars were included in this analysis.  Of these,
151,722 were
observed throughout all 3 quarters.  Other targets were observed during a
subset of quarters due to positioning (some stars are observable in some quarters
but not others due to small alignment errors in the focal plane's CCD detectors), 
and changes in the target selection algorithm over time.
There were 2,659 target stars observed in quarter 1 only; 1,723 stars in quarter 2 only;
1,394 stars in quarter 3 only; 
1,085 stars in quarters 1 and 2; 11,671 stars in quarters 2 and 3;
and 690 stars in quarters 1 and 3.  The detections cataloged in Table \ref{t1} indicate,
for each target, the quarters during which photometric data was acquired for that target.

\subsection{Pre-Search Processing}\label{presearch}

The pixel data from the spacecraft are processed by the \kepler{} photometry pipeline
prior to the search for transiting planet signals.
The first processing step is performed by the Calibration (CAL) 
module \citep{evq2010}.  Calibration includes detection and removal of cosmic rays in 
collateral (non-science) pixels, correction for
variations in pixel sensitivity, removal of detector bias and dark current
effects, removal of smear artifacts introduced by shutterless operation, and
removal of distortions introduced by the readout electronics.  Following
calibration, the Photometric Analysis (PA) module 
\citep{jdt2010a}             
assembles the data from
individual pixels into flux time series for each target.  This includes performing
aperture phototometry on the calibrated pixels associated with each target, 
estimation and removal of backgrounds, and detection and removal of cosmic rays
in science pixels.  The
flux time series from PA are then processed by the Pre-Search Data Conditioning
(PDC) module 
\citep{jdt2010b}.  
The PDC module performs a number of
error corrections upon each flux time series.  The corrections include compensation
for effects which are systematic across the focal plane, such as changes in flux due to
pointing variation, differential velocity aberration, focus changes, and thermal transients;
there are also corrections which are specific to each target, such as 
removal of step discontinuities
due to cosmic ray damage to individual CCD pixels and 
removal of excess flux due to aperture crowding.  Finally, PDC fills gaps introduced
by intra-quarter spacecraft anomalies or operations.

The processing steps described above are performed on a quarterly
basis, and result in piecewise-continuous flux time series with inter-quarter
gaps that is presented to the TPS module. The data conditioning
steps described above are intended to remove all sources of flux variation with the 
exception of astrophysical phenomena and uncorrelated instrument and statistical
noise.  

\subsection{Post-Search Analysis}

The procedure for discovery of extrasolar planets does not end with the detections
produced by the Transiting Planet Search (TPS) algorithm as described in Section 
\ref{tps}.  The detections from TPS are further analyzed by the Data Validation
(DV) module \citep{hw2010}.  DV performs a number of automated tests which assist
in evaluating the probability that a TPS detection is an extrasolar planet.  The
results of the DV tests are collected and presented to the \kepler{} science team.

The Data Validation module is the final automated step in the discovery process.  All
subsequent steps are labor-intensive and performed by members of the science team.
The results from DV are examined and assessed, detections which are obviously
not due to astrophysical phenomena are eliminated, and the remainder are classified
as Kepler Objects of Interest (KOIs) \citep{wjb2011}.  Further analyisis of the KOIs
permits them to be sorted into one of several categories:  likely planet detections;
eclipsing binary stars; false positives (typically background eclipsing binaries); 
and other detections which evade classification.  The likely planet detections are
prioritized and subjected to follow-up observations using a wide variety of ground-based
and space-based instruments.  The follow-up observations will permit validation of
some of the likely planet detections, while others will be revealed as additional
false positive detections.  

As the foregoing description shows, it is not possible to classify TPS detections
without considerable subsequent analysis and additional measurements from other
instruments.  In the remainder of this paper we will consider the full complement of TPS
detections in the Q1-Q3 dataset; as the majority of these have not yet received
the necessary ``person-in-the-loop'' analysis, it is not possible to say with any certainty which
ones are likely to be signals from extrasolar planets and which ones are likely to 
be some form of false positive\footnote{The exception to this is TPS detections which 
correspond to known KOIs, as discussed in Section \ref{koi}.}. It is anticipated that 
the presented information on the full set of detections will
assist researchers by guiding them to target stars in the \kepler{} field of view which
are thought to be of the greatest interest from the viewpoint of transiting objects.
The table can also be used as a catalogue of known detections, so that future detections
of candidates by other researchers using other techniques can determine whether their
detection corroborates a known event or represents a potentially heretofore-unknown event.

\section{Transiting Planet Search}\label{tps}

The Transiting Planet Search (TPS) module is described in some detail elsewhere
\citep{jmj2002,jmj2010}, and we briefly summarize its main features and actions
here.

As described in Section \ref{presearch}, the processing steps prior to TPS invocation
all operate on data from one quarter at a time.  Thus, the first task of TPS is
to combine the flux time series from several quarters into a single flux time
series suitable for searching.  This ``quarter stitching'' procedure
is as follows:%
\begin{itemize}
\item The data from each quarter is converted from absolute flux in photoelectrons
   to fractional flux variation; this is accomplished by dividing each quarter's flux
   by that quarter's median flux value and subtracting 1
\item strongly sinusoidal variations are removed from the flux quarter-by-quarter:
   the variations are first identified by constructing a periodogram of the flux,
   and identifying well-separated peaks which are strong compared to an estimate of
   the noise at the selected frequency (the latter being determined from the power
   spectral density in the region surrounding each peak); the harmonics are then
   fitted via Levenberg-Marquardt algorithm, with the amplitude, phase, and frequency
   of each harmonic as fit parameters; the fitted harmonics are then subtracted from
   the flux time series
\item Any trends present in each quarter at the start and end of data acquisition are
   removed:  a third order polynomial is constructed which matches the average offset
   and slope of the flux time series in its first and last 49 cadences; subtracting
   this polynomial removes the net offset and slope from the ends of the time series,
   at the expense of introducing potentially large excursions to the remainder of the
   time series; these excursions are then removed by fitting a polynomial of the form
   $x(1-x)p(x)$, where x is normalized time (i.e., $x\equiv0$ at the start of the
   quarter and $x\equiv1$ at the end) and $p(x)$ is a conventional polynomial of the
   form $p(x) = p_0 + p_1x + p_2x^2 + ...$; this process reduces the excursions while
   preserving the desired edge conditioning from the initial cubic polynomial subtraction;
   the polynomial order is determined through use of Akaike's Information Criterion 
   \citep{ha1974}, subject to a user-selected limit (set to 10th order in this case)
\item Remaining data gaps are filled:  this includes inter-quarter gaps and
   gaps which are a full quarter or longer due to incomplete observations for some
   targets; note that this is done in order to facilitate construction of a digital filter
   bank which is required for the subsequent analysis.  The gap-filling algorithm
   is the same as the one used in PDC 
   \citep{jdt2010b}.  A detailed description of the algorithm is beyond the scope
   of this paper, and in any event studies have indicated that the detection of
   transits depends only weakly on the gap-filling algorithm.
\end{itemize}

The second step in the TPS analysis process is to perform a detection of individual
transits in the quarter-stitched flux time series.  This process is complicated by the
fact that the noise due to stellar variability is both non-white and 
non-stationary, indicating that a joint time-frequency analysis procedure will be
required.  The procedure is described in detail in \citet{jmj2002}, and is summarized
in Appendix \ref{wavelet}.  This results in a Single Event Statistic (SES) time series:
a time series which describes, for each sample in the original flux time series, 
the significance (in $\sigma$) of the detection of a transit-like signal centered on
that sample.  As shown in Appendix \ref{wavelet}, the algorithm
is sensitive to the match in overall shape between the transits in the data and the 
model transit pulse used in the search; thus, the detection signal will be maximized
when the duration of the latter is well-matched to the duration of the former, and
falls off when the two are mismatched in duration.  The search is therefore performed
repeatedly using a discrete set of different model transit durations in order to
ensure sensitivity to all expected transit durations, and a Single Event Statistic
time series is produced for each model transit duration used in the search.  
The number of transit durations used in the search is set by balancing the
conflicting demands of maximum sensitivity to a wide variety of transit durations (which
mandates a large number of searches, with small increments in trial transit pulse duration
from one search to another) and computational tractability (which mandates a small number
of searches).
Additionally, the selected model transit durations must be comparable to the 
expected actual durations of planetary transits.  Based on these requirements, 
each flux time series is searched for individual transit-like features with the following
14 durations:  1.5, 2.0, 2.5, 3.0, 3.5, 4.5, 5.0, 6.0, 7.5, 9.0, 10.5, 12.0, 
12.5, and 15.0 hours.  

The next step is to use the results of the search for individual transit-like features
to search for transit-like features which are periodic.  This is accomplished
by combining, or ``folding'' the Single Event Statistics across time into Multiple
Event Statistics (MES) according to Equation \ref{mes}.  
The periods which are used in each folding are a compromise between small period 
spacing to minimize false negatives and large period spacing to minimize computing
requirements \citep{jmj1996}.  
The set of periods is determined algorithmically during execution using specified
values for the minimum and maximum periods and the desired model correlations between
consecutive periods; for the purposes of this analysis, the minimum period used in
folding was 0.5 days, and the maximum was the full duration of the dataset, or 218 
days.  At each period, the search examines a set of possible phases, where the phase
step is determined from the duration of the trial transit pulse in such a way as to
provide a user-selected degree of correlation from one phase step to the next.
The
result of folding is a 2-dimensional array of Multiple Event Statistics, one value for
each combination of period and phase.  These are collapsed along the phase axis to yield
a vector of Multiple Event Statistics versus period, where the vector value represents the
maximum of the Multiple Event Statistics versus phase for that particular period.  It is
at this point that gap-fills are suppressed from contributing to a detection:  when a
filled cadence is included in a fold, it is not permitted to contribute to the
Multiple Event Statistic.  

It is interesting to note that, while the process above is explicitly designed to find
transit-like features which are periodic, the TPS algorithm has had reasonable
success in detecting planet candidates which exhibit transit timing variations (TTVs)
which cause their transit timings to depart from perfect periodicity.  This is because
TPS uses a dense grid of periods; if the individual transit signals are sufficiently
large and the transit timing variations sufficiently small,
detections can occur when the ``wings'' of the transits are folded over one
another.  Examples of this include Kepler-9b and -9c \citep{mjh2010}, 
and all of the Kepler-11 planets \citep{jjl2011}.

\begin{figure}
\plotone{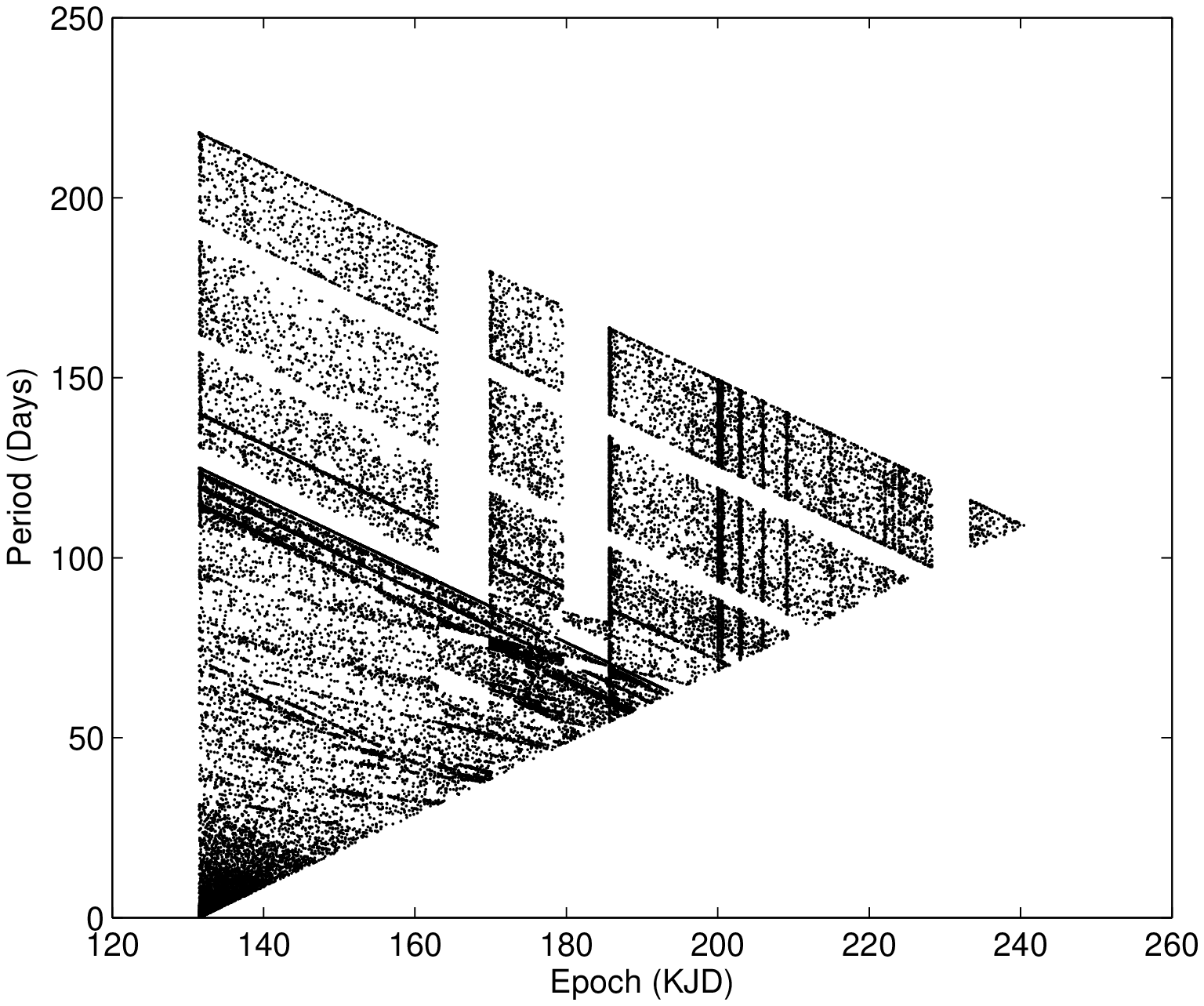}
\caption{Epochs and periods for all events which pass the Multiple
Event Statistic criterion.  The structures in this feature correspond to known
spacecraft events:  vertical features to events in the first half of the Q1-Q3
interval, diagonal features to events in the second half of the interval.
\label{f1}}
\end{figure}

The final step is to perform cuts based on the results of the above analyses.  The first
and most important cut is to remove from further analysis any star for which all of the Multiple Event Statistics for all periods and all pulse durations falls below 
$7.1\sigma$.  The threshold value of $7.1\sigma$ was chosen as a compromise which
simultaneously limits the number of expected false positive detections due to pure
statistical fluctuations, while on the other hand preserving sensitivity to 
Earth-analogs \citep{jmj2002b}.
For the dataset spanning the first 3 quarters, the
number of stars which have at least one Multiple Event Statistic of at least $7.1\sigma$
is 66,593.  Thus, this cut eliminates 104,261 targets from further analysis, 
or 61\%.  Figure \ref{f1} shows the period of each transit in days, and the epoch
of each transit in Julian Date offset by 2,454,833 days (so-called ``Kepler-modified 
Julian Date'', or KJD).  The strong features shown in Figure \ref{f1} are caused by
a variety of known spacecraft events:  reaction wheel desaturations, small adjustments
in spacecraft attitude during data acquisition, and thermal transients due to the
large change in spacecraft attitude required for monthly data downlinks.  

\begin{figure}
\plotone{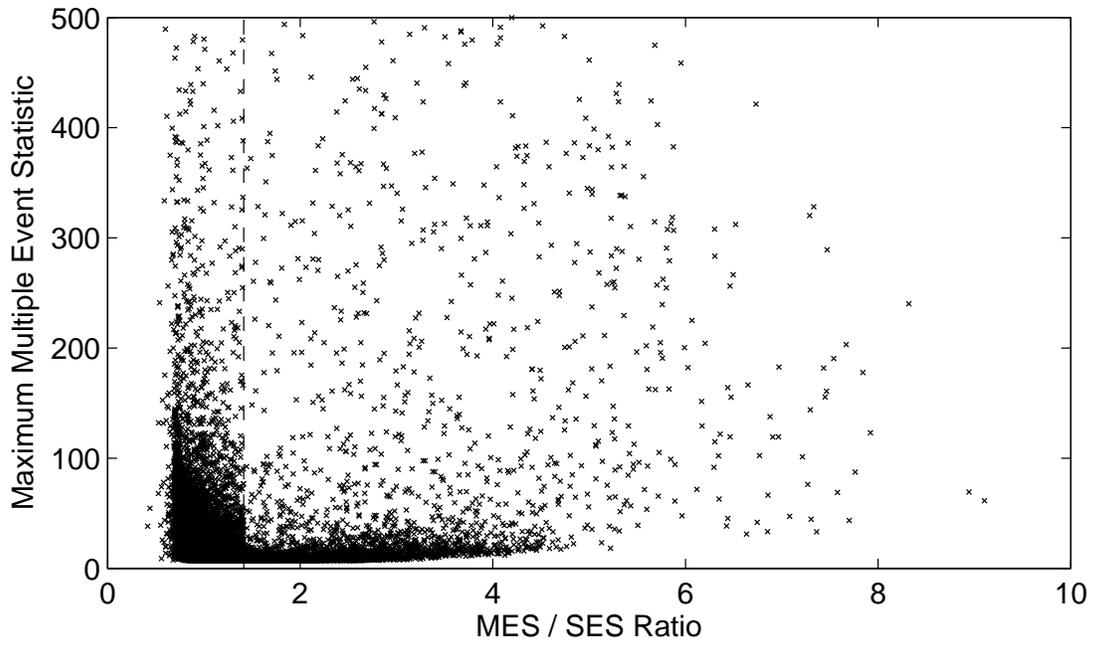}
\caption{Multiple Event Statistics and MES/SES
 ratios for targets for which the Multiple Event Statistic exceeds
 $7.1\sigma$.  The dashed line is at MES/SES $\equiv\sqrt{2}$.  \label{f2}}
\end{figure}

The second cut makes use of the relationship between the maximum Multiple Event
Statistic and the several Single Event Statistics which were combined to yield that
Multiple Event Statistic.  In particular, a threshold is established for the 
rato of the Multiple
Event Statistic and the largest participating Single Event Statistic.  As Figure
\ref{f2} shows, there are two distinct populations:  a population with a low
MES/SES ratio, and a population with larger values of this ratio.  The dividing line
is at a MES/SES of approximately $\sqrt{2}$.  The qualitative explanation for this
distinction is as follows:  in the case of a signal with 2 transits, in which the
transits are of equal depth and occur against backgrounds of comparable noise, the 
ratio of MES/SES
will be $\sqrt{2}$, and this ratio rises approximately as the square root of the 
number of transits, as shown in Equation \ref{mes};
when the ratio of MES/SES is lower than $\sqrt{2}$, this indicates that
two highly-unequal Single Event Statistics have been combined into a Multiple Event
Statistic, a situation which is unlikely to occur in the case of 
transits of uniform depth and thus is unlikely to represent a true transit signal.  
For this reason, we apply a cut on the MES/SES ratio
of 1.4142.  This cut eliminates a large number of false positives which combine 
highly unequal Single Event Statistics into a Multiple Event Statistic, but also
eliminates true positives which contain 2 transits of near-equal depth.
Thus the MES/SES cut is also an
implicit cut on orbital period:  a planet which has a period greater than half the 
duration of the dataset will produce only 2 transits within the dataset, which will
yield a MES/SES which is below the cutoff.  The
number of target stars which satisfy both the Multiple Event Statistic and the
MES/SES criteria is 5,392.

\section{Detected Signals of Potential Transiting Planets}

\begin{figure}
\plotone{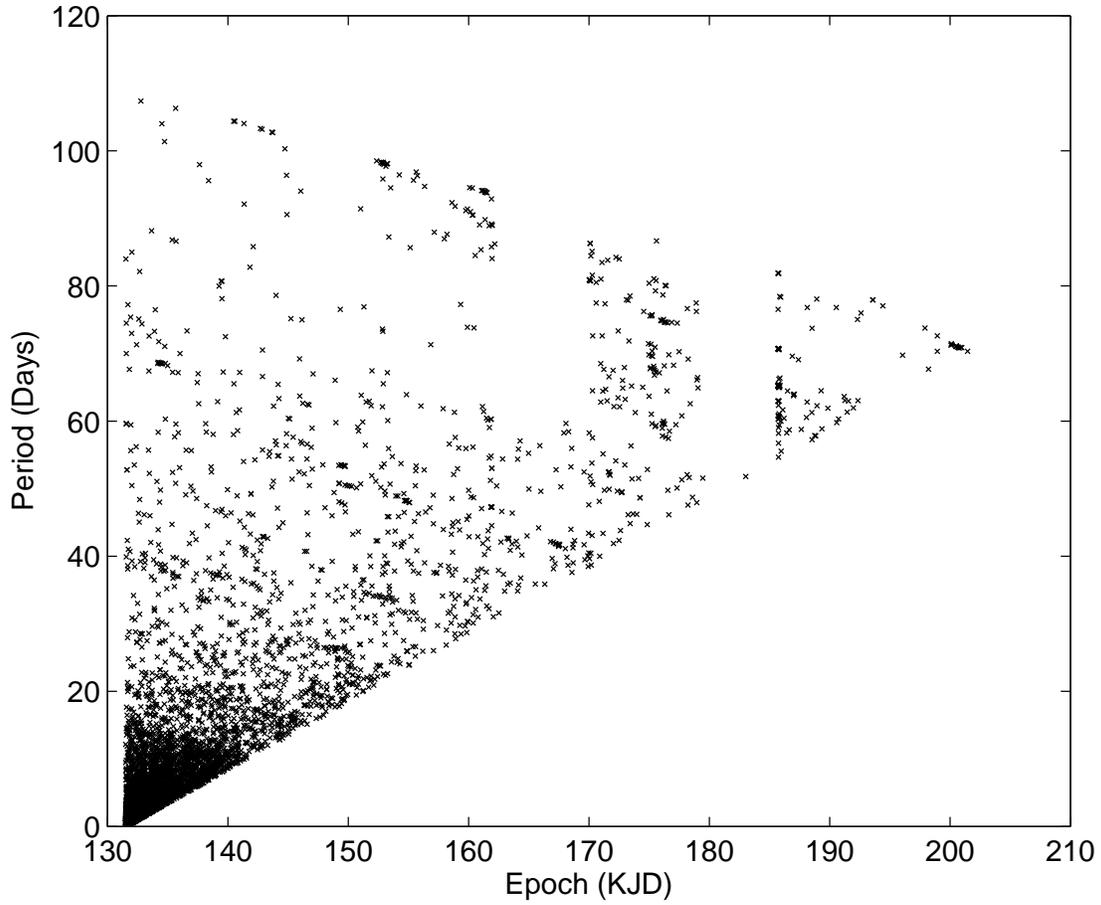}
\caption{Epochs and periods for detected signals.  The regions
in parameter space which contain no detections are due to
gaps in data acquisition combined with the requirement that a detection 
achieve a MES/SES ratio of at least $\sqrt{2}$.    Note the
change in scale compared to Figure \ref{f1}, demonstrating that the MES/SES cut
does effectively reduce the range of periods which can be detected when that cut
is enforced.
\label{f3}}
\end{figure}

Figure \ref{f3} shows the epoch 
and period of the 5,392 TPS detections.  Note that, as compared with
Figure \ref{f1}, Figure \ref{f3} is devoid of strong features from spacecraft events;
in addition, a number of spurious detections unrelated to spacecraft events are
eliminated.  However, the range of periods in Figure \ref{f3} is about half that of
those in Figure \ref{f1}, illustrating the implicit period limitation of the MES/SES
cut.
Figure \ref{f3} shows a relatively large number of detections with short periods, 
which is expected for two reasons:  first, short-period signals will be relatively easier to detect
than long-period signals of equal depth due to the larger number of transits within
the Q1-Q3 interval;  second, the probability of a planetary system's inclination 
angle being such that the transits are visible to \kepler{} is inversely 
proportional to the semi-major axis of the planet's orbit.  
Figure \ref{f3} also shows a small number of regions of
parameter space which contain no detections.  These
are due to data gaps.  For example, consider the region centered on an
epoch of KJD 166 and periods of approximately 80 days.  A signal with a period and
an epoch in this region would have only 3 transits in the data interval
which is presented here, one of which would fall within the long interval between
quarter 1 and quarter 2 data acquisition; the remaining 2 transits would not
pass the MES/SES ratio cut; therefore, no signal which meets the
\kepler{} detection criteria can have such a combination of parameters.  

The most striking feature of Figure \ref{f3} is the 3 hard edges which bound all of
the detections.  The left-hand edge is due to the start of observations, since by
definition no detection can have an epoch of first transit which precedes the start
of observations.  No detection can occur below the lower edge because any such
detection has a period and epoch such that it would have a transit which occurs at
an earlier epoch, which lies within the populated region (for example, a detection
which has an Epoch of KJD 150 and period 10 days is impossible, because such a detection
would also have a transit at KJD 140, and so would be represented by a point at KJD
140 and period 10 days).  The upper edge represents detections with 3 transits for which
the third transit occurs in the last few cadences of Q3; a detection which lies above
this edge is impossible because its third transit occurs after the end of Q3, and the
remaining 2 transits would be eliminated by the MES/SES cut.

\begin{figure}
\plotone{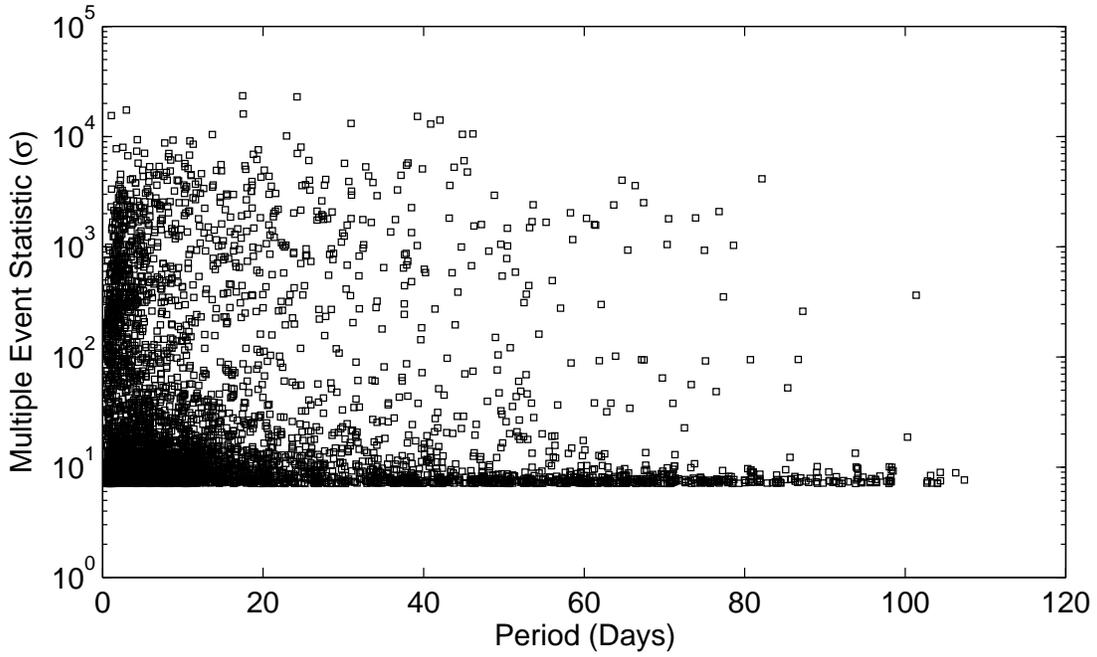}
\caption{Distribution of periods and multiple event statistics
for detected signals.  The hard edge in the distribution is at 7.1 $\sigma$, the
threshold for a detection.\label{f4}}
\end{figure}

Figure \ref{f4} shows the detections as a function of period and
multiple event statistic.  As with Figure \ref{f3}, there is a gradient in detections
which favors shorter periods over longer ones, and weaker signals over stronger ones.  

\begin{figure}
\plotone{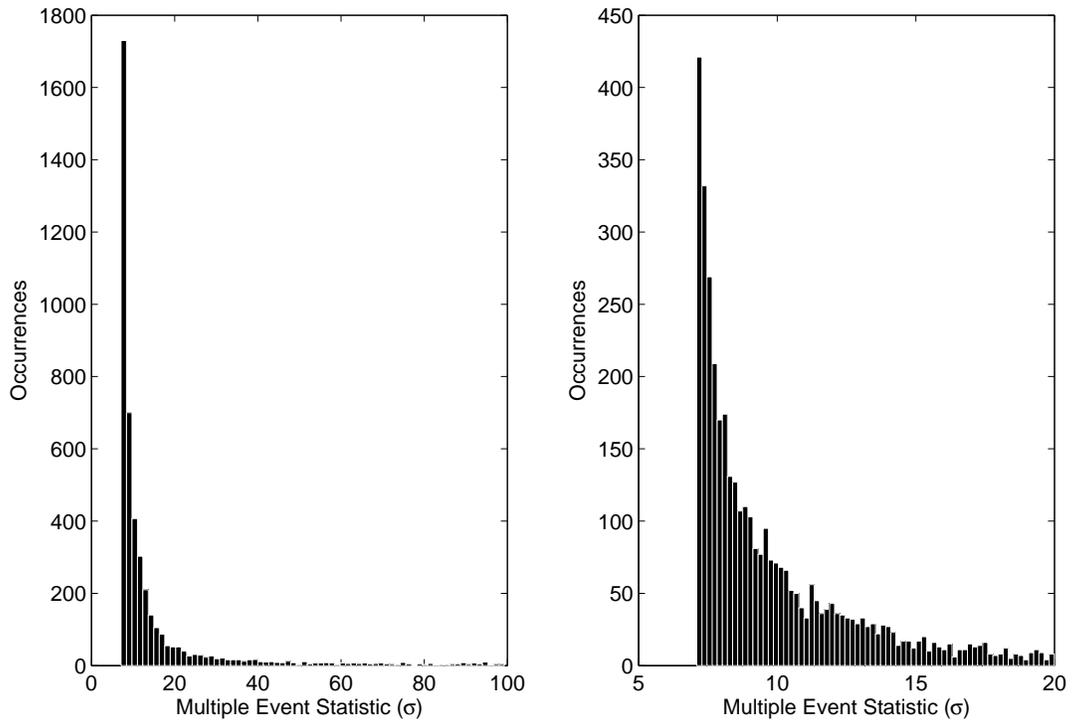}
\caption{Histogram of maximum Multiple Event Statistics.  Left:  4,424 out
of 5,392 detections with maximum Multiple Event Statistic under $100\sigma$.
Right:  3,780 out of 5,392 detections with maximum Multiple Event Statistic
under $20\sigma$.
\label{f5}}
\end{figure}

\begin{figure}
\plotone{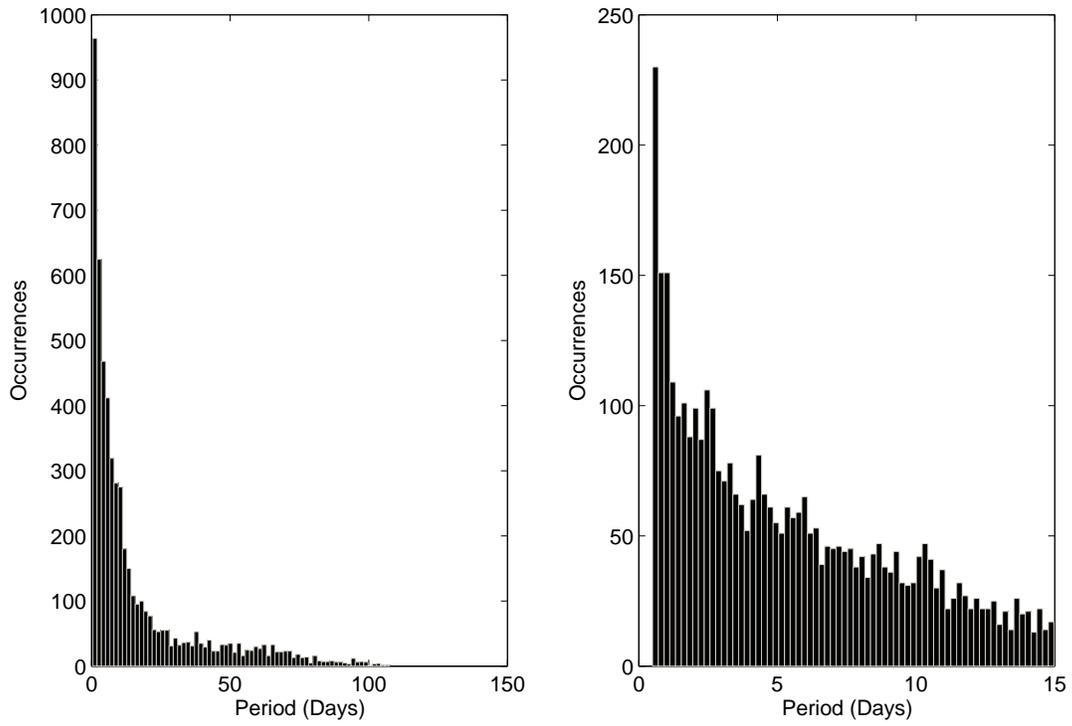}
\caption{Histogram of periods of detected signals.  Left:  All 5,392 detections.
Right:  3,732 out of 5,392 detections with period under 15 days.  The hard edge
at 0.5 days is due to the selection of this as the minimum search period for the
study.
\label{f6}}
\end{figure}

Figures \ref{f5} and \ref{f6} quantify the degree to which the distribution of detected
signals is skewed towards weak signals and short periods.  In Figure \ref{f5} we see the
distribution of Multiple Event Statistics.  Although the detections include signals as
strong as $23,520\sigma$, there are 4,424 detections out of 5,392 which have a maximum
Multiple Event Statistic under 100 $\sigma$, as shown in the left panel of Figure \ref{f5}, and
3,780 detections which have a maximum Multiple Event Statistic under 20 $\sigma$, as shown in
the right panel of Figure \ref{f5}.  Similarly, the left side of Figure \ref{f6} shows
the period distribution of all detections; the right side of Figure \ref{f6} shows the
period distribution of the 3,732 detections which have periods under 15 days.
It is interesting to note that, based on Figures \ref{f4}, \ref{f5} and \ref{f6}, 
the number of
detections in the first three quarters which could plausibly be detected using a single
quarter's worth of data is about 2,400; the remaining 3,000 detections possess periods 
too long or signals too weak to be detected using only a single quarter of data.

Although TPS does not directly compute the transit depth of each detection, it is
possible to crudely estimate the transit depth.  
For each detection, the period and the epoch are used to identify the cadences which
fall closest to the center of each transit; the noise characteristics at those cadences
and the Multiple Event Statistic can be combined to yield an estimate of the transit
depth.  Because the trial transit pulse used by TPS is square rather than transit-shaped, 
and in general will not
be precisely matched to the actual duration of the detected transit signature, 
the resulting estimate of transit
depth (in parts per million, or PPM) is probably only reliable at the factor-of-2 level,
but it does give a general sense of whether the dataset is dominated by 
deep or shallow transits.

\begin{figure}
\plotone{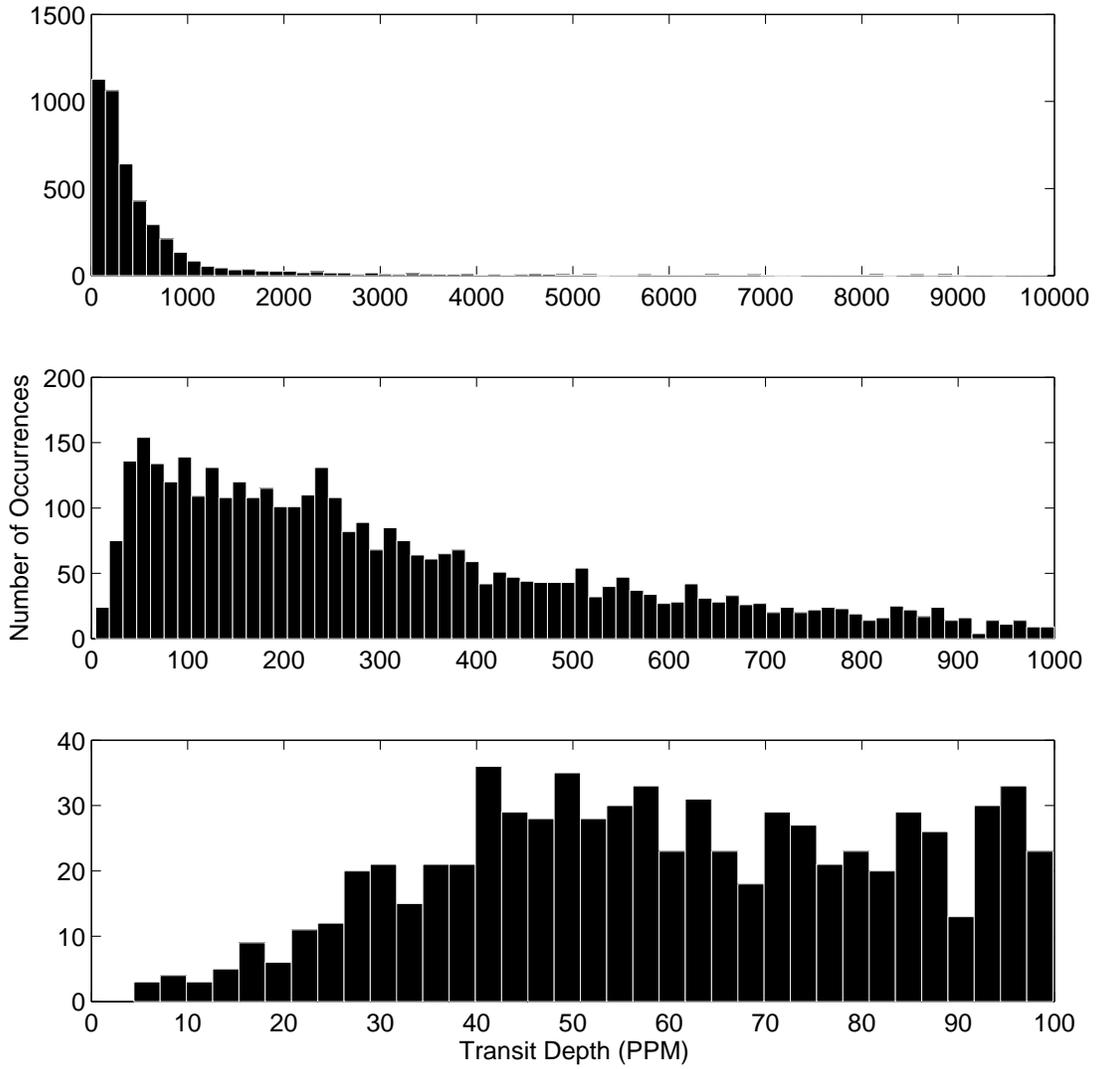}
\caption{Transit depths.  Top:  4,581 detected signals with transit
depths up to 10,000 PPM.  Middle:  3,900 detected signals with transit depths up to
1,000 PPM.  Bottom:  739 detected signals with transit depths up to 100 PPM.
\label{f7}}
\end{figure}

\begin{figure}
\plotone{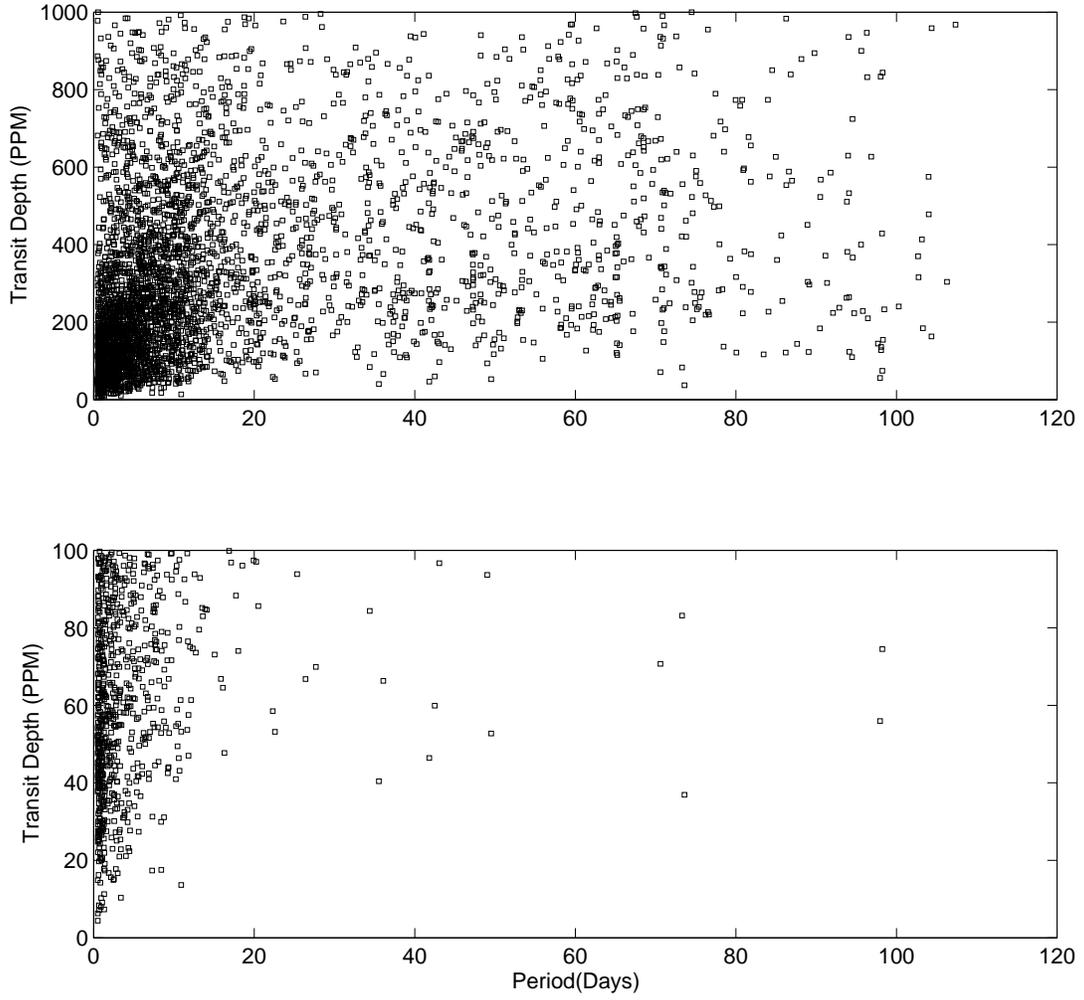}
\caption{Transit depths and orbital periods.  Top:  3,900 detected
signals with transit depths up to 1000 PPM.  Bottom:  739 detected signals with
transit depths up to 100 PPM.
\label{f8}}
\end{figure}

Figure \ref{f7} shows the distribution of transit depths over three ranges:  transit
depths up to 1\% (10,000 PPM), transit depths up to 0.1\% (1000 PPM), and transit depths
up to 0.01\% (100 PPM).  There are 4,581, 3,900, and 739 targets, respectively,
in these distributions.  The 712 targets which exhibit transit depths greater than
10,000 PPM form a long, diffuse tail which extends up to 415,000 PPM; these objects
include a few super-Jovian planets and a large number of eclipsing binaries.
Figure \ref{f8}
shows the relationship between transit depth and period.  It is 
both apparent and unsurprising that the
smallest transits detected correspond to the shortest periods.

Of the 5,392 targets with detected signals, 3,815 of those targets were observed in all
three quarters.  Twelve detections were on targets observed in quarters 1 and 2, 192
on targets observed in quarters 2 and 3, 12 on targets observed on quarters 1 and 3;
1,035 on targets observed in quarter 1 only, 291 on targets observed in quarter 2 only,
and 35 on targets observed in quarter 3 only.  The large number of detections on
targets observed only in Q1 is possibly explained by the fact that most of the targets which
were removed from the observation list after Q1 were red giants.  The oscillations
which are a characteristic of red giants produce false-positive detections
in TPS:  these oscillations are in an unfortunate amplitude regime, in which they
are too weak to trigger the harmonic removal procedure described in Section \ref{tps}, 
but are still strong enough to trigger detection by the main TPS algorithm.  
It is entirely possible that some of the detections on targets observed
only in Q1 are actual transiting planet signatures, but at this time the TPS algorithm
has no means to distinguish between a true detection and this form of false positive.

Table \ref{t1} shows a complete list of detections in the first 3 quarters of \kepler{}
data.  The printed version shows only the first 90 detections,
while the online version shows the full 5,392.  Table \ref{t1} includes all of the
detections on stars observed only in Q1, though this set of detections is thought to be
dominated by false positives.  The uncertainty shown for transit depth is the statistical
uncertainty, which is determined by the signal-to-noise of the detection, and does
not include the systematic biases due to mismatches between the model transit and
the actual transit in shape and duration, as described above.

\subsection{Example Target Data Sets and Processing}

\begin{figure}
\plotone{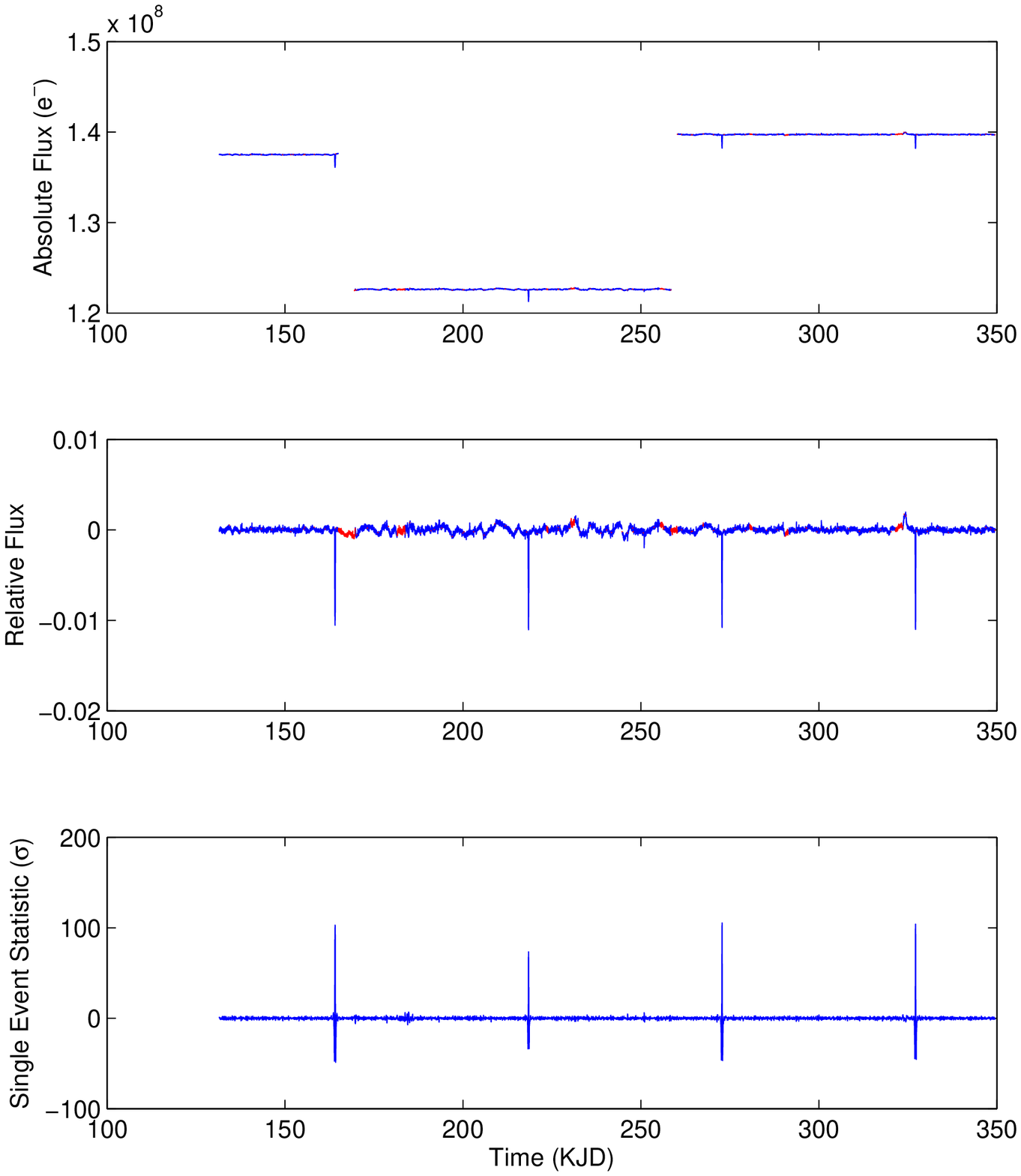}
\caption{Illustrated analysis procedure for KIC 2309719.  Top:  initial flux
time series presented to TPS, with gap-fill values in red and data values in blue.
Middle:  quarter-stitched flux time series, with gap-fill values in red and data
values in blue.  Bottom:  Single Event Statistics time series for detection of a
transit with 3.5 hour duration.
\label{f9}}
\end{figure}

\begin{figure}[t]
\plotone{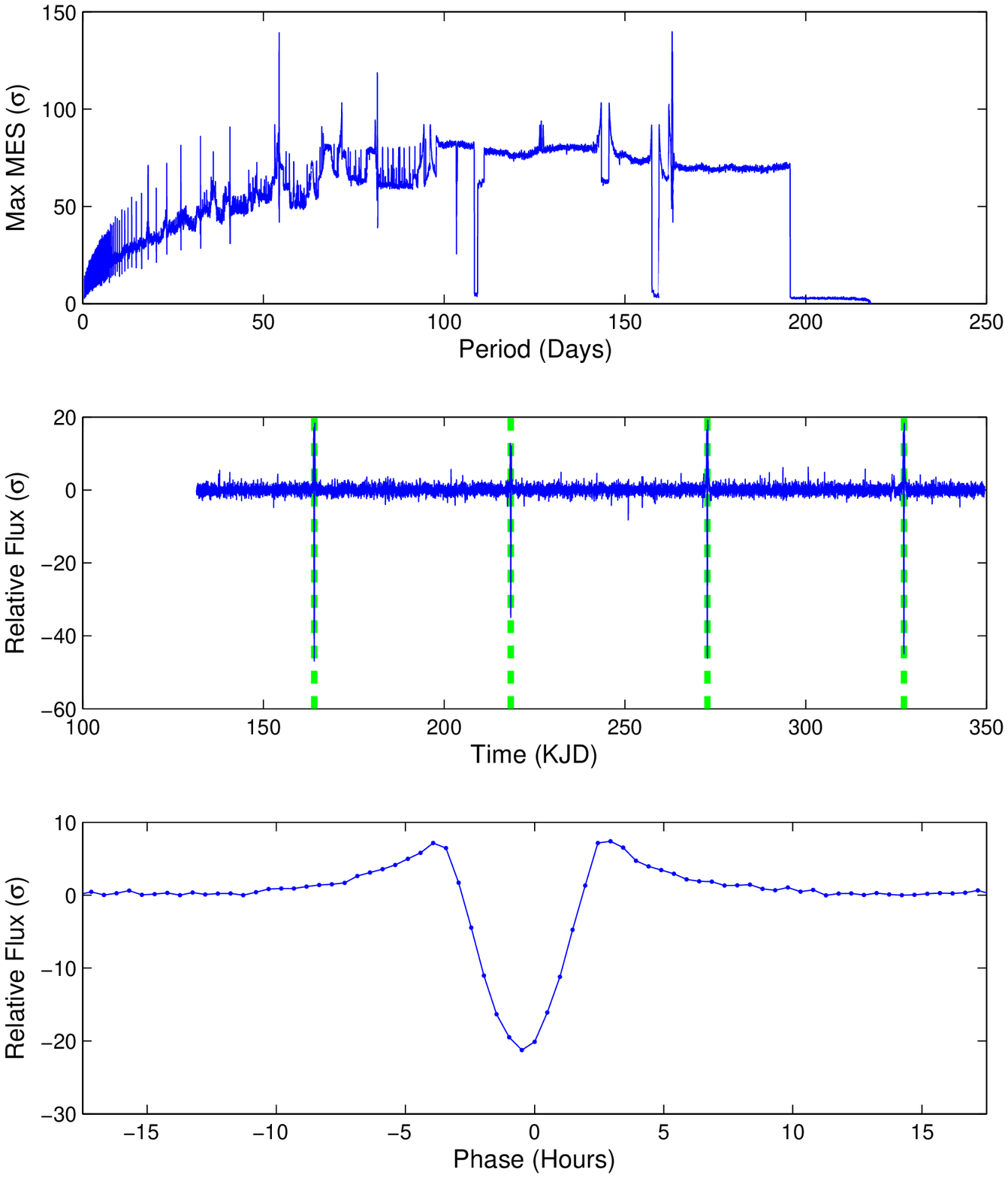}
\caption{Illustrated analysis procedure for KIC 2309719, continued.  Top:
maximum Multiple Event Statistic as a function of period for detection of
transits with 3.5 hour durations.  
Middle:  whitened flux
time series, with expected transit times indicated by green, dashed vertical lines.  Bottom:
whitened flux time series folded at the detection period of 54.36 days, summed into
bins of 29.4 minute width, and zoomed on the detection phase.
\label{f10}}
\end{figure}

\begin{figure}
\plotone{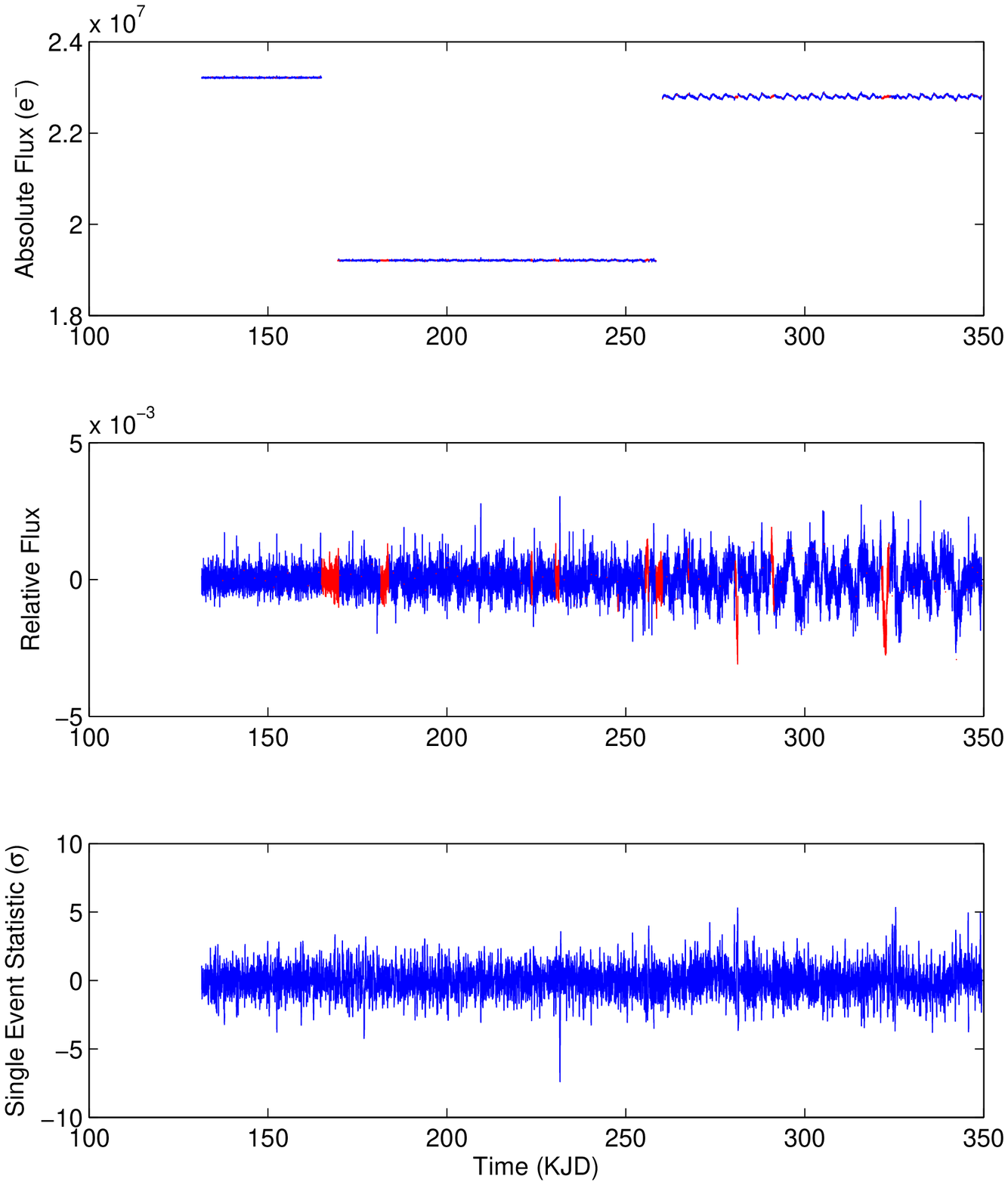}
\caption{Illustrated analysis procedure for KIC 2010191.  Top:  initial flux
time series presented to TPS, with gap-fill values in red and data values in blue.
Middle:  quarter-stitched flux time series, with gap-fill values in red and data
values in blue.  Bottom:  Single Event Statistics time series for detection of a
transit with 2.5 hour duration.
\label{f11}}
\end{figure}

\begin{figure}
\plotone{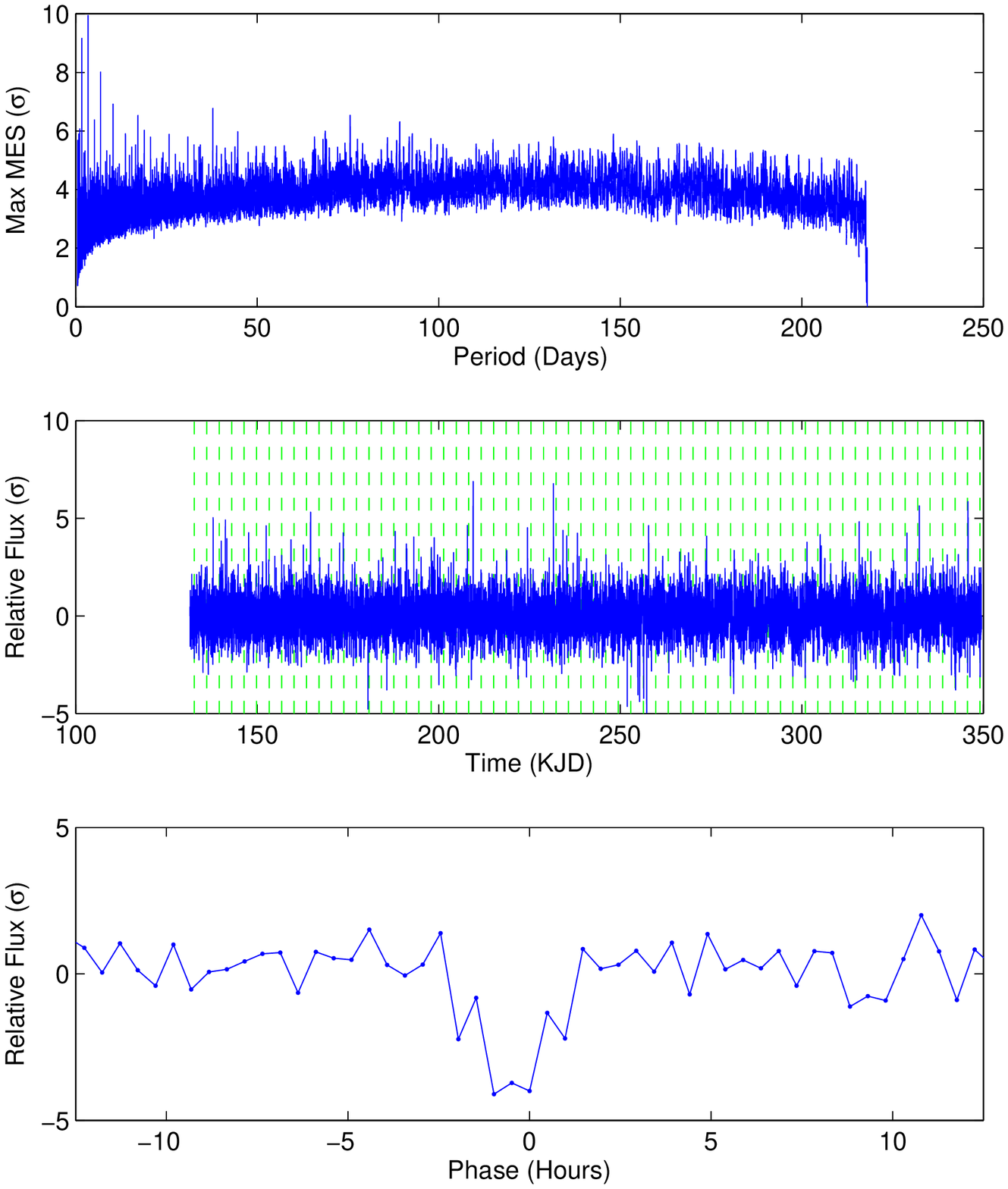}
\caption{Illustrated analysis procedure for KIC 2010191, continued.  Top:
maximum Multiple Event Statistic as a function of period for detection of transits
with 2.5 hour durations.  
Middle:  whitened flux
time series, with expected transit times indicated by green, dashed vertical lines.  Bottom:
whitened flux time series folded at the detection period of 3.43 days, summed into
bins of 29.4 minute width, and zoomed on the detection phase.
\label{f12}}
\end{figure}

Figures \ref{f9}, \ref{f10}, \ref{f11}, and \ref{f12} show examples of \kepler{} light
curves, and the process for analyzing the light curves and detecting the signals of
transiting planets.

Figures \ref{f9} and \ref{f10} show the analysis of Kepler Input Catalog (KIC) 2309719, 
a star with Kepler Magnitude 12.9 that was observed in all 3 quarters.  This target contains 
a transit signature which is large enough to be seen with the unaided eye:  a 1.1\%
depth with a period of 54.36 days.  The top plot of Figure \ref{f9} shows the 
original flux time series, with intra-quarter gap-filled values in red and data
values in blue.  The middle plot of Figure \ref{f9} shows the quarter-stitched flux
time series; in this case, both intra- and inter-quarter gap filled values are in 
red.  The bottom plot of Figure \ref{f9} shows the Single Event Statistics time series
for this target, using a trial transit duration of 3.5 hours.  
In all three time series, the signature of the repeating transit
event is easily detected by observation.

The top plot of Figure \ref{f10} shows the maximum Multiple Event Statistic as a 
function of period, again using a trial transit duration of 3.5 hours.  
The peak at 54.36 days is clearly visible, as is a slightly 
lower peak at three times the actual period.  The peaks at periods shorter than
54.36 days are at periods which are harmonically related to the actual period:
for example, folding the data at a period of 40.77 days will fold the first and
last actual transits on top of one another, resulting in an increase in MES relative
to periods which cause the first and last transits to ``miss'' one another in
folding.  Similarly, there is a continuum of values which are above the 7.1 $\sigma$
threshold because a single transit can have a significance as large as 100 $\sigma$,
as shown in Figure \ref{f9}; thus, folding a single transit on top of several
noise-dominated samples can easily exceed 7.1 $\sigma$.  
The middle plot of Figure \ref{f10}
shows the whitened flux time series for KIC 2309719:  note that the structure
which is evident in Quarter 2 of the quarter-stitched original flux, in the middle
plot of Figure \ref{f9}, has been removed by the whitener, and that the shape of the
transits has been distorted, with positive excursions occurring prior to and 
preceding the transit; these positive excursions in the whitened flux time series
result in negative excursions in the Single Event Statistics time series in Figure 
\ref{f9}.  
Green, dashed vertical lines in this plot 
indicate the expected transit 
locations based on the period and epoch of the detection, and as expected they
coincide with the visible transit signatures.
Finally, the bottom plot of Figure \ref{f10} shows the 
whitened flux time series after it has been folded at the detection period of 54.36
days, collected into 29.4 minute (1 long cadence exposure) wide time bins, summed
within bins, rescaled to unit variance, and zoomed in upon the phase of the transit.
As expected, when processed in this manner the transit signature is clearly visible
to casual observation.

Figures \ref{f11} and \ref{f12} show the same series of plots which follow the
processing of KIC 2010191, a star with Kepler Magnitude 14.6 which was
observed in all 3 quarters.  In this case, the maximum Multiple Event statistic
occurs for a trial transit duration of 2.5 hours.
This target contains a transit signature which is too
small to be discerned with the unaided eye, with a depth of approximately 285 parts
per million and 3.43 day period.  Although the plot of Maximum Multiple Event
Statistic (top plot of Figure \ref{f12}) clearly shows a detection at the 3.43 day
period, the event only becomes apparent in the folded,
binned, and summed version of the whitened flux time series, in the bottom of 
Figure \ref{f12}.

\subsection{Comparison with Known Kepler Objects of Interest (KOIs)}\label{koi}

The optimal means of assessing the reliability of TPS detections is to introduce into
TPS a set of flux time series which contains a known population of transit-like events,
and compare the resulting detections to that known population.  Under current
circumstances this is not possible, since the only source of flux time series which 
sufficiently resemble the real behavior of stars is the actual stars in the \kepler{} field of
view, and determining which of these stars have transiting planets is the goal of
the mission -- the unknown which we seek to make known.  Nonetheless, at this time it
is possible to perform a partial test of TPS by comparing the TPS detections to the
current list of Kepler Objects of Interest (KOIs) \citep{wjb2011}.  The KOI list is a set
of transit detections which have been investigated in detail by the \kepler{} Science Team using
a variety of techniques, and represents our best current knowledge of astrophysical signals
in the \kepler{} field of view.  Importantly, none of the KOIs were detected using the 
multi-quarter TPS algorithm described above, but were instead detected using a variety of
alternate methods, including use of the TPS algorithm on individual quarters of data.  
While the KOI list is obviously not yet a complete catalog of every
astrophysical transit-like signature in the \kepler{} field of view, by determining the 
fraction of KOIs which are detected by TPS we can gain insight into the reliability of
the TPS algorithm overall, which has implications for the reliability of TPS detections which
are not on the KOI list.

The current list of Kepler Objects of Interest
contains 1,235 planet candidates, plus 498 objects which are either eclipsing binaries
(foreground or background) or are a transiting planet on a background object
(``false positives'').  For the purposes of this analysis, the aggregate of the 
planet candidate list and the false positive list shall be referred to as the ``KOI list.''
The list was compiled using the first five quarters of \kepler{} data, which has two
consequences for the current comparison.  First, any KOI which has fewer than three transits
in the Q1-Q3 interval cannot be detected by TPS; there are 139 such KOIs, which we have
exluded from the comparison.  Second, the KOI detections will have a more favorable
signal to noise ratio than the Q1-Q3 TPS detections, which will result in some of the
latter falling below the detection threshold.  No attempt has been made to identify or
remove the targets which fall into this category.  In
addition, TPS returns only the most significant detection in a time series, and so any
target which contains more than one KOI must be reduced to a single KOI for 
purposes of this comparison\footnote{The full analysis pipeline includes algorithms 
which allow detection of multiple
transiting planet candidates on a single star, for details see \citep{hw2010,pt2010}.}.
In such cases, it is necessary to determine which KOI on a given target corresponds to
the TPS detection, if any.  This is accomplished by comparing the period of each KOI on a
given target star with the period of the detection, and selecting the KOI which 
most closely agrees in period with
the detection, taking into account the fact that the TPS period can 
differ from the KOI period by a
rational factor (2, 0.5, 3, 0.333, etc.).  Removal of multiple KOIs per target star eliminates
221 KOIs from the comparison, yielding 1,373 KOIs which we expect to be potentially 
detectible within the Q1-Q3 dataset.  

\begin{figure}
\plotone{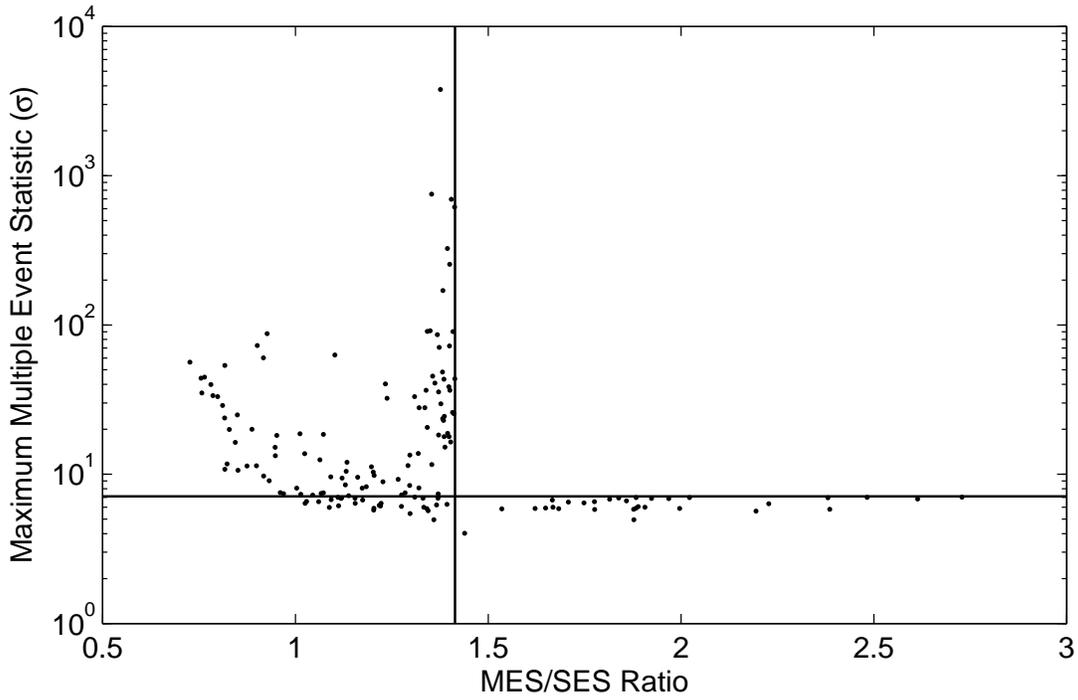}
\caption{Multiple Event Statistic and MES/SES ratio for target stars which contain a
Kepler Object of Interest (KOI), but did not meet the criteria for a detection in the Q1-Q3 dataset.
The vertical line indicates the cut on MES/SES ratio, the horizontal line indicates
the cut on Multiple Event Statistic value.  The upper-right quadrant represents events
which pass both cuts, and is thus empty.
\label{f13}}
\end{figure}

Out of the 1,373 KOIs described above, only 1,242 (90.5\%) meet both the Multiple Event 
Statistic and MES/SES ratio criteria for a detection in TPS.  Figure \ref{f13} shows
the Multiple Event Statistic and MES/SES ratio for each target star which
contains a KOI but which did not meet both the criteria required to be considered
as a detection in the Q1-Q3 dataset.  Due to the aforementioned SNR consideration, 
it is not a surprise that some KOIs were not detected due to insufficient signal;
similarly, the fact that longer flux time series, with more transits, were used in
construction of the KOI list makes it inevitable that some KOIs were rejected in the
current analysis due to low MES/SES ratio.
The remainder of this analysis will be confined to the 1,242 target stars
which present both a KOI and a detection.

\begin{figure}
\plotone{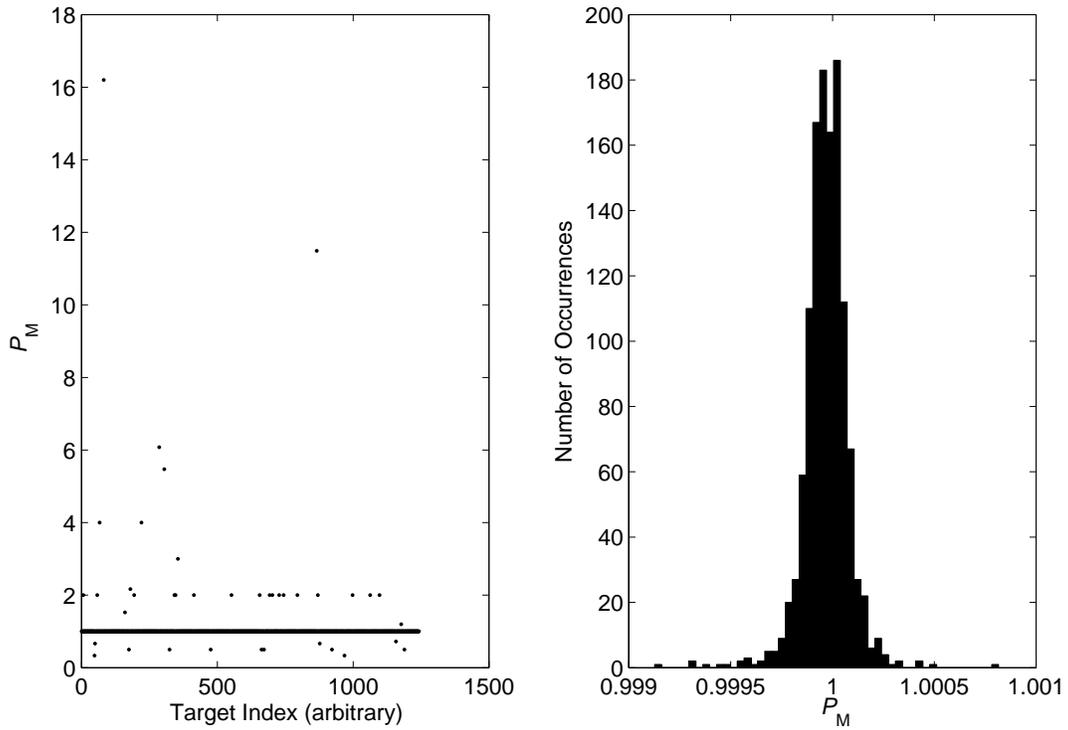}
\caption{Left:  all values of the period figure of merit, $P_{\rm M}$, as defined in
the text.  Right:  Distribution of $P_{\rm M}$ values clustered about 1.0.  Note that
97.4\% of all values (1,210 out of 1,242) have $P_{\rm M}$ equal to 1.0 or a nearby
rational number.
\label{f14}}
\end{figure}

The most straightforward way to determine whether the TPS detections agree with the 
KOI list is to compare the epochs and periods of the two.  More specifically, it is
possible to define a period figure of merit, $P_{\rm M}$, which is the ratio of the
KOI period to the TPS period for a given target star; and an epoch figure of merit,
$E_{\rm M}$, which is the difference between the KOI and TPS epochs divided by the KOI
period\footnote{In fact, what is used is the corrected KOI epoch:  the earliest
expected transit time of the given target star in the Q1-Q3 period.  Use of this
epoch removes an ambiguity in the KOI epoch, specifically that the KOI epoch can
differ from the TPS epoch by an integer number of periods.}.  Figure \ref{f14} shows
$P_{\rm M}$:  the left plot shows all the values of $P_{\rm M}$, while the right
plot shows the distribution of values clustered around $P_{\rm M}=1.0$.  Note that the
distribution in the right-hand plot is quite narrow, with a standard deviation
of $1.10\times10^{-4}$.  The number of targets for which $P_{M}$ lies
within 0.001 of 0.2, 0.25, 0.333, 0.5, 1.0, 2.0, 3.0, or 4.0 is 1,232 out of 1,242, 
or 99.1\%.

\begin{figure}
\plotone{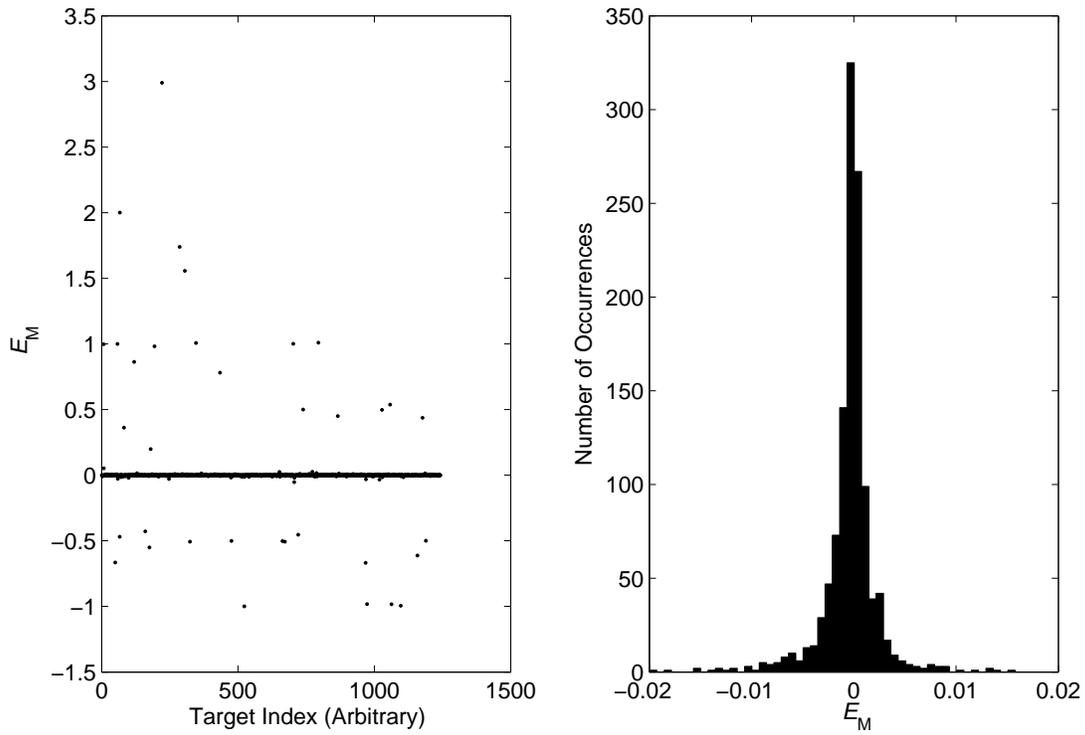}
\caption{Left:  all values of the epoch figure of merit, $E_{\rm M}$, as defined in
the text.  Right:  Distribution of $E_{\rm M}$ values clustered about 1.0.  Note that
97.3\% of all values (1,209 out of 1,242) have $E_{\rm M}$ approximately equal to -1.0, 
0, 1.0, or 2.0.
\label{f15}}
\end{figure}

Figure \ref{f15} shows, on the left, all values of $E_{\rm M}$.  The distribution of
$E_{\rm M}$ is not as tight as for $P_{\rm M}$, as shown in the right hand plot
of Figure \ref{f15}:  the cluster around $E_{\rm M}=0$ has a standard deviation of
$2.12\times10^{-3}$.  
The total number of targets which have $E_{\rm M}$ within 0.02 of
-1, 0, +1, +2, or +3 is 1,210 out of 1,242, or 97.4\%.

The number of target stars which have values of both $E_{\rm M}$ and 
$P_{\rm M}$ which indicate agreement between the KOI list and the TPS detections
is 1,209 out of 1,242, or 97.3\%. 
Each of the remaining possible combinations
of $E_{\rm M}$ and $P_{\rm M}$ values is discussed below.

\subsubsection{Agreement in Period but not Epoch}

A total of 23 target stars agree in period between the KOI list and the TPS
detections, but not in epoch.  In this case, ``agreement'' can include periods
which differ by a rational factor, as described above.  Of these, 14 are eclipsing
binaries, and their epoch disagreements fall into two categories:  in 11 cases, the 
TPS epoch is for the primary eclipse and the KOI is for the secondary, or vice-versa;
in 3 cases, the target has a continuous, near-sinusoidal oscillation at the eclipse
period, which can introduce an irresolvable ambiguity in the TPS epoch determination.
Of the remaining targets with good period agreement but a discrepancy in epoch, 2
targets appear to disagree by a time which is very close to 1 long cadence interval of
29.4 minutes; two targets were ``overfolded,'' yielding a TPS period which is a
fraction of the KOI period (1/2 and 1/3, respectively) and a TPS epoch which is
displaced from the KOI epoch by 1 and 2 TPS periods, respectively; 1 target was in
transit at the start of observations, causing an edge-effect confusion.  The remaining
4 targets have unexplained offsets in epoch, which are all on the order of two
long cadence intervals.  

\subsubsection{Agreement in Epoch but not Period}

Only 1 target star agrees in epoch but disagrees in period between the KOI list and 
the TPS detections.  In this case, the TPS and KOI periods are rationally related, but
in an unexpected way:  the TPS period is 2/3 of the KOI period (0.79815 days versus
1.1973 days, respectively).  

\subsubsection{Agreement in Neither Period nor Epoch}

A total of 9 target stars show disagreement in both period and epoch between the
KOI list and the TPS detections.  There is no clear unifying feature to their flux
time series, but in general they appear to be poorly conditioned:
large variations which have been poorly corrected in PDC, significant numbers of
negative outlier values, etc.  It is likely that all 9 of these cases contain artifacts
or stellar variations which have been misidentified by TPS, and the fact that these
target stars also contain KOIs is due to random chance.  

\subsubsection{Summary of Comparison}

Out of the 1,373 KOIs which have 3 or more transits in the Q1-Q3 \kepler{} dataset, a total of
1,209 were correctly detected by TPS, although in a few cases the TPS-detected period
is a harmonic or subharmonic of the KOI period, and in some other cases the TPS-detected
epoch is similarly displaced relative to the KOI epoch.  This represents a correct 
detection rate of 88.1\%.  Of the balance, the largest effect is false negatives -- target
stars for which no TPS detection was reported.  False negatives account for 131 KOIs,
or 9.5\% of the KOIs.  
These cases are believed to be primarily due to the fact that the TPS detections are
based upon Q1-Q3 data while the KOI list is based upon Q1-Q5 data; the additional quarters
of data approximately double the number of observations,
resulting in an expected 40\% increase in signal-to-noise.
An additional 23 targets (1.7\%) demonstrated agreement in period but not epoch, though the causes
of most of these discrepancies are understood.  An
additional 1 target star (0.07\%) apparently detected the KOI-specified epoch but not the
KOI-specified period nor any harmonic or subharmonic of same. Finally, 9 stars (0.7\%) reported a 
TPS detection
which does not agree with either the epoch or the period of the corresponding KOI; these
cases are thought to be false positive detections which occured on KOI targets solely because of
random chance.

\section{Future Plans and Open Research Areas}\label{future}

The 5,392 detected signals in this dataset are not expected to include every true
positive transiting planet signal, nor are they expected to be free of false 
positive signals.  Improving the performance of the TPS algorithms on both of these
topics is an active area of development.  

The most likely cause of missed signals is the cut on MES/SES.  From Figure \ref{f13},
we can see that
there are a number of transiting planet candidates in the KOI catalog with 
MES/SES below the
current threshold of $\sqrt{2}$, which indicates that the hard cut currently in use
is causing false negatives in some cases.  Additionally, it seems
likely that the hard cut is causing false negatives in long-period cases in which
only 2 transits are expected in a given dataset.  Development effort is focused on
finding a means to eliminate this somewhat heuristic cut, which would probably 
require implementation of a more suitable algorithmic approach that would eliminate
the large number of false positives which are currently removed by the MES/SES cut
while preserving the smaller number of true positive detections.

One promising candidate for improving both false positive rejection and true
positive retention is implementation of a robust statistic.  The current TPS
algorithm performs a rather simple combination of Single Event Statistics into a
Multiple Event Statistic.  This is a computationally tractable approach, but it
means that TPS is not sensitive to cases in which single events of wildly different
depth are combined into a Multiple Event Statistic (i.e., TPS has no means to 
reject a detection which ``does not look like a transit'').  The robust statistic
approach would follow the initial TPS detection with a robust fit of a periodic
square pulse to the light curve.  This has the potential to reject detections which
do not have the characteristic near-constant depth of a real transit while
preserving detections which are currently eliminated by the MES/SES cut.

The \kepler{} target list includes several thousand eclipsing binary stars.
Because of the detailed structure of these light curves, it is difficult for
TPS and the downstream automated processing system to function correctly on these; this
is most especially the case for contact binaries.  As a consequence, it is rarely
the case that a system with an eclipsing binary can be searched for a transiting 
planet as well.  In the future, this will be addressed by making use of our
existing prior knowledge of the list of eclipsing binary target stars to subject
these targets to an additional pre-processing step which removes the eclipsing
binary signature and simplifies the task of searching for planets.

At this time, there is no algorithm in the \kepler{} analysis pipeline which 
records the presence of single, non-periodic events which are potential transits.
In some cases, such single events are potentially interesting in that they may
represent a true transiting object for which the second transit has not yet occurred.
These signals would potentially serve as an indicator of target stars which are 
deserving of further scrutiny, as a second transit may occur later in the mission.
Even if no such transit occurs, a single deep transit could represent an extrasolar
planet with a long period, for which the second transit will not occur 
during the lifetime of the \kepler{} spacecraft.  

As mentioned above, the TPS algorithm is designed to detect transits which are 
perfectly periodic, and is capable of detecting transits with small amounts of
TTV, albeit with reduced significance.  It is likely that transiting planet
signals with TTV are being missed, either because the detection statistic is
reduced sufficiently by TTV to fall below the TPS threshold, or because the
TTVs are so large that no detection can occur.  Development of a transit detection
algorithm which can accommodate larger TTVs and remains computationally affordable
is an active area of research on the \kepler{} Mission.

Finally, in addition to all the projected improvements to TPS itself, all of the
other software modules in the \kepler{} analysis pipeline are also undergoing
continual improvement.  It is expected that improvements in the CAL, PA, and PDC
modules will result in the presentation of higher-quality flux time series to TPS,
and a corresponding improvement in planetary detection and false-positive rejection
relative to current capabilities.

\acknowledgments

Funding for this mission is provided by NASA's Space Mission Directorate.  The
contributions of Hema Chandrasekaran, Chris Burke, Jennifer Hall,
Khadeejah Ibrahim, and Kamal Uddin have been essential in the studies documented
here.



{\it Facilities:} \facility{Kepler}.



\appendix

\section{Joint Time-Frequency Analysis of Flux Time Series}\label{wavelet}

The problem of transit detection can be expressed mathematically as the
search of a flux time series, $x(n)$, for a particular signal, $s(n)$, where
$x(n)$ contains a non-white, non-stationary, and initially unknown noise
spectrum.  For purposes of this discussion, and in the Transiting Planet
Search algorithm, $s(n)$ is given by a square pulse of a selected duration,
but more complex shapes can also be used.

The first step in the process is to decompose $x(n)$ into its frequency
components in a manner which preserves the time variation of the noise.
This is done by constructing a digital filter bank, using Daubechies'
12-tap wavelet \citep{daub1988}; wavelets are used because of their
duration-limited nature, which permits preservation of time variation
within a selected frequency band.  The filter bank is constructed in octaves,
such that each output of the filter bank is centered on a frequency which is
exactly half that of the next-higher output and exactly twice that of the 
next-lower output.  The output of the filter bank is a
set of time series $x_1, x_2 ... x_M$, where $x_1$ contains the frequency
content of $x$ near the Nyquist frequency, $x_2$ contains the content centered
on half of the Nyquist frequency, etc., until $x_M$, which contains the remaining
low frequency content of the original flux time series, including the DC term.
The same filter bank is used to decompose the signal $s$ into a set of time
series $s_i$.

The time-varying noise in each frequency band can be computed from the
variance of the $x_i$ time series:
\begin{equation}\label{sigma}
\hat{\sigma}_i^2(n) = \frac{1}{2^iK+1} \sum_{n-2^{i-1}K}^{n+2^{i-1}K} x_i^2(k),
\end{equation}
where $K$ determines the window over which the variance is computed.  The
value of $K$ must be chosen such that it is long compared to the typical periods
in $x_1$, and also such that it is long compared to the duration of a transit.
In practice, TPS uses a value of $K$ which is 30 times the duration of the
model transit.  Note that, for transits which are small compared to the overall
variation of the initial time series, the frequency content of the transit is a
small perturbation on the overall frequency content, and so Equation \ref{sigma}
is valid.  For large transits, for example Jovian planets or eclipsing binaries,
the transit itself can distort the estimate in Equation \ref{sigma}.  For this
reason, in practice TPS uses a Median Absolute Deviation (MAD) rather than a
true variance, and scales the MAD by the inverse MAD of a unit-width Gaussian.
This gives a value for $\hat{\sigma}_i$ which is not distorted by the presence
of large transits, so long as the duty cycle of the transits is small.

With $x_i$ and $\hat{\sigma}_i$ time series computed, it is possible to perform
pointwise division of the former by the latter.  The resulting time series,
$x_i / \hat{\sigma}_i$, is equal to the contribution from the
$i$th frequency band to white Gaussian noise with unit variance, plus a 
perturbation from the transit signals (if any are present).
In principle, we can now correlate the ``whitened'' time series 
$x_i / \hat{\sigma}_i$ with the time series $s_i$ to determine the strength
of the latter signal in the former time series, and we can combine the resulting
correlations to determine the total transit signal as a function of time within
the flux time series.  However, the pointwise-division of the vector $x_i$
by the noise vector $\hat{\sigma}_i$ has introduced a time-varying distortion
to the shape of the transit, and this distortion must be incorporated into the
correlation process by applying the same distortion to the signal vector $s_i$.
By rearranging terms, this can be expressed as a ``double whitening'' of $x_i$,
which permits us to write an expression for the correlation between $x_i$ and
$s_i$:
\begin{equation}\label{correlation1}
\mathbb{N}_i(n) = \left\{\left[\frac{x_i}{\hat{\sigma^2_i}}\right]*\tilde{s}_i\right\}(n), 
\end{equation}
where $\tilde{s}_i$ is the time-reversal of $s_i$, and $*$ represents convolution.
$\mathbb{N}_i$ represents the contribution from frequency band $i$ to the detection
of $s$ in $x$.  The contributions from all bands can be summed, with appropriate
scaling coefficients to take account of the fact that the transform used here is 
{\it overcomplete} (i.e., the number of samples in each $x_i$ is equal to the number
of samples in $x$, thus the number of samples across all $x_i$ is $M$ times as large
as the number of samples in $x$, and this over-representation must be corrected).  The
resulting sum is the contribution from all bands to the detection of $s$ in $x$:
\begin{equation}\label{correlation2}
\mathbb{N}(n) = \sum_{i=1}^{M} 2^{-\min(i,M-1)} \mathbb{N}_i.
\end{equation}

While Equation \ref{correlation2} allows determination of the amplitude of the transit
detection as a function of sample number, the significance of this detection can only
be determined by comparing that detection amplitude to the detection amplitude which is
expected from the noise spectrum.  This can be determined in a manner analogous to the
detection amplitude, and in the interest of simplicity we simply reproduce the result
from \citet{jmj2002}:
\begin{equation}\label{normalization}
\mathbb{D}(n) =  \sum_{i=1}^{M} 2^{-\min(i,M-1)} 
\left[\hat{\sigma}_i^{-2} * \tilde{s}_i^2\right](n).
\end{equation}
Equation \ref{normalization} can be obtained from Equation \ref{correlation1} by
substituting $\hat{\sigma}_i$ for $x_i$, and by combining the contributions from 
individual bands in quadrature, rather than linearly.  This corresponds to an
intuitive understanding of what it means to estimate the expected detection amplitude
from stellar noise sources.

With the definitions above, we can define a time series which represents the detection
significance, in $\sigma$, as a function of sample number. This is obtained by
pointwise-division of $\mathbb{N}$ by the square root of $\mathbb{D}$:
\begin{equation}
l(n) \equiv \frac{\mathbb{N}(n)}{\sqrt{\mathbb{D}(n)}}.
\end{equation}
The time series $l(n)$ is the {\it Single Event Statistics} time series.

The detection significance of a transit series can be obtained by combining the
$\mathbb{N}$ and $\mathbb{D}$ values at the locations of the individual transits.
The {\it Multiple Event Statistic} for detection of a sequence of transits at
widely spaced locations is given by:
\begin{equation}\label{mes}
l \equiv \frac{\sum_k\mathbb{N}(k)}{\sqrt{\sum_k\mathbb{D}(k)}},
\end{equation}
where $k$ runs over the locations of the prospective transits in the time series.
In the limit where all of the $\mathbb{N}$ values are equal to one another, and all
of the $\mathbb{D}$ values are also equal to one another, Equation \ref{mes} reduces
to the familiar increase in detection significance as the square root of the number
of events.





\clearpage

\clearpage

\clearpage

\clearpage

\clearpage

\clearpage

\clearpage

\clearpage

\clearpage

\clearpage

\clearpage

\clearpage

\clearpage

\clearpage

\clearpage

\clearpage

\begin{deluxetable}{ccccrrrr}
\tablecaption{List of detections in first 3 quarters of \kepler{} data.
As noted in the text, the transit depth values are an approximation.
Published KOIs are indicated by a number in the KOI column; 
published false positives are indicated by ``FP' in the KOI column.
\label{t1}}
\tablewidth{0pt}
\tablehead{
\colhead{Kepler} & \colhead{Kepler} & 
\colhead{Quarters} & \colhead{KOI or} & \colhead{Epoch} & 
\colhead{Period} & \colhead{MES} & 
\colhead{Depth} \\ 
\colhead{ID} & \colhead{Magnitude} & \colhead{Observed} &
\colhead{FP} &
\colhead{(KJD)} & \colhead{(Days)} & \colhead{$\sigma$} & 
\colhead{(PPM)}
}
\startdata
757450 & 15.3 & 123 & 889 & 134.44 &   8.89 &    128.4 &  12624.0 $\pm$    98.4 \\ 
1026032 & 14.8 & 123 &  & 133.77 &   8.46 &    663.5 &  53537.7 $\pm$    80.7 \\ 
1026473 & 13.8 & 123 &  & 141.82 &  82.78 &      7.2 &   1379.7 $\pm$   192.2 \\ 
1026957 & 12.6 & 123 & FP & 144.77 &  21.76 &     20.8 &    582.3 $\pm$    28.0 \\ 
1161345 & 11.6 & 123 & 984 & 133.34 &   4.29 &     48.0 &    534.5 $\pm$    11.1 \\ 
1295531 & 11.9 & 123 &  & 131.57 &   0.70 &      7.3 &    133.0 $\pm$    18.1 \\ 
1430444 & 13.5 & 123 &  & 170.04 &  39.62 &      7.5 &    281.2 $\pm$    37.4 \\ 
1430741 & 13.9 & 123 &  & 131.51 &   0.71 &      8.2 &    118.6 $\pm$    14.4 \\ 
1431122 & 14.6 & 123 & 994 & 132.84 &   4.30 &      7.4 &    243.4 $\pm$    32.9 \\ 
1432047 & 11.0 & 1 &  & 135.82 &  10.56 &     10.8 &    395.0 $\pm$    36.4 \\ 
1433899 & 15.2 & 123 &  & 131.74 &   0.52 &      7.5 &    120.0 $\pm$    15.9 \\ 
1434259 & 12.2 & 123 &  & 185.70 &  81.88 &      8.6 &    375.4 $\pm$    43.4 \\ 
1569823 & 10.6 & 1 &  & 138.83 &   9.28 &      7.9 &    161.0 $\pm$    20.3 \\ 
1571088 & 10.4 & 123 &  & 152.86 &  95.82 &      7.9 &    228.8 $\pm$    29.0 \\ 
1571511 & 13.4 & 123 & FP & 135.51 &  14.02 &    623.0 &  14997.8 $\pm$    24.1 \\ 
1572201 &  8.6 & 123 &  & 149.79 &  35.54 &      8.9 &     40.4 $\pm$     4.6 \\ 
1572768 &  9.5 & 1 &  & 135.59 &   5.37 &      9.2 &    235.1 $\pm$    25.5 \\ 
1573954 & 14.8 & 123 &  & 175.06 &  75.56 &      7.6 &    559.1 $\pm$    73.4 \\ 
1717242 & 14.7 & 2 &  & 144.34 &  26.66 &      7.2 &    298.4 $\pm$    41.4 \\ 
1724968 & 13.4 & 123 &  & 160.15 &  90.98 &      7.2 &  12591.1 $\pm$  1736.8 \\ 
1725193 & 14.5 & 23 &  & 134.11 &   2.96 &    383.0 &  15701.1 $\pm$    41.0 \\ 
1725415 & 14.5 & 123 &  & 131.93 &   2.44 &      7.3 &     93.7 $\pm$    12.9 \\ 
1849195 & 16.7 & 2 &  & 137.03 &  12.23 &      7.6 &   1222.1 $\pm$   160.0 \\ 
1865042 & 13.6 & 123 & 1002 & 132.98 &   3.48 &      8.0 &    123.9 $\pm$    15.4 \\ 
1865448 & 12.4 & 1 &  & 131.79 &   2.04 &      8.9 &    133.8 $\pm$    15.0 \\ 
1865910 & 10.5 & 1 &  & 133.98 &   5.63 &     11.3 &    266.8 $\pm$    23.7 \\ 
1871056 & 13.0 & 123 & 1001 & 155.68 &  40.82 &     12.2 &    274.4 $\pm$    22.5 \\ 
1871465 & 12.1 & 123 &  & 134.15 &  68.53 &      7.3 &    642.2 $\pm$    88.4 \\ 
1872166 & 11.6 & 123 &  & 185.77 &  66.20 &      7.7 &    733.9 $\pm$    94.9 \\ 
1872948 & 10.5 & 123 &  & 131.72 &   1.39 &      7.8 &     17.4 $\pm$     2.2 \\ 
1873157 & 13.7 & 123 &  & 178.92 &  76.23 &      7.6 &    527.4 $\pm$    69.4 \\ 
1995489 & 12.2 & 1 &  & 132.07 &   1.23 &      8.0 &     49.2 $\pm$     6.2 \\ 
1995732 & 15.7 & 23 &  & 171.32 &  77.36 &    350.9 &  57996.4 $\pm$   165.3 \\ 
1996180 & 13.9 & 123 &  & 134.35 &   3.03 &      7.4 &    130.1 $\pm$    17.6 \\ 
2010191 & 14.6 & 123 &  & 132.66 &   3.43 &     10.0 &    285.3 $\pm$    28.6 \\ 
2010607 & 11.3 & 123 &  & 141.57 &  18.64 &     25.6 &    329.7 $\pm$    12.9 \\ 
2012722 & 11.7 & 123 &  & 132.08 &   0.78 &      7.5 &     29.2 $\pm$     3.9 \\ 
2013502 & 11.9 & 123 &  & 185.75 &  65.14 &      8.3 &    872.8 $\pm$   104.6 \\ 
2014991 & 12.4 & 123 &  & 132.18 &   6.50 &      8.2 &    136.6 $\pm$    16.7 \\ 
2015864 & 13.1 & 1 &  & 138.36 &   6.90 &      8.1 &    163.6 $\pm$    20.2 \\ 
2018112 & 11.8 & 1 &  & 132.81 &   2.79 &     14.7 &    341.0 $\pm$    23.3 \\ 
2019076 & 15.7 & 23 &  & 131.85 &   3.56 &    438.4 &  49476.3 $\pm$   112.9 \\ 
2020966 & 12.1 & 1 &  & 132.78 &   4.88 &      8.2 &    299.7 $\pm$    36.7 \\ 
2021440 & 15.6 & 123 &  & 136.32 &   7.81 &      9.2 &    562.6 $\pm$    60.9 \\ 
2021736 & 14.4 & 23 &  & 150.26 &  50.33 &      8.3 &    691.3 $\pm$    83.6 \\ 
2140491 & 10.2 & 2 &  & 142.73 &  12.06 &      8.7 &    139.2 $\pm$    16.0 \\ 
2140780 & 10.8 & 1 &  & 136.31 &   5.97 &     10.7 &    241.4 $\pm$    22.6 \\ 
2141387 & 12.2 & 123 &  & 185.95 &  55.54 &      7.4 &   5566.3 $\pm$   750.8 \\ 
2142628 & 11.7 & 1 &  & 135.28 &  11.40 &      8.0 &    273.4 $\pm$    34.1 \\ 
2157247 & 14.4 & 123 & FP & 133.31 &   5.69 &     13.1 &    547.3 $\pm$    41.8 \\ 
2162994 & 14.2 & 123 &  & 132.63 &   2.05 &   1121.6 &  72694.6 $\pm$    64.8 \\ 
2163644 & 15.1 & 123 &  & 170.12 &  80.75 &      7.8 &   1332.8 $\pm$   171.8 \\ 
2164169 & 14.8 & 123 & 1029 & 133.87 &  32.31 &      9.2 &    637.8 $\pm$    69.2 \\ 
2165002 & 15.4 & 123 & 999 & 146.16 &  16.57 &     10.8 &   1070.1 $\pm$    99.1 \\ 
2166206 & 13.3 & 123 & FP & 132.63 &   8.10 &     25.5 &    557.1 $\pm$    21.9 \\ 
2166534 & 15.4 & 123 &  & 177.01 &  68.26 &      8.4 &   1384.6 $\pm$   164.7 \\ 
2166962 & 15.0 & 123 &  & 170.05 &  80.90 &      8.9 &   1219.6 $\pm$   136.4 \\ 
2282763 & 12.5 & 123 &  & 146.22 &  62.64 &      7.4 &    958.3 $\pm$   130.3 \\ 
2284079 & 10.4 & 1 &  & 132.22 &   2.48 &     11.8 &    224.0 $\pm$    19.0 \\ 
2284344 & 11.6 & 1 &  & 132.38 &   4.11 &      8.6 &     65.7 $\pm$     7.6 \\ 
2300399 & 12.4 & 1 &  & 133.98 &   7.40 &     11.4 &    331.0 $\pm$    28.9 \\ 
2300529 & 12.6 & 123 &  & 131.65 &   0.51 &      8.0 &    113.3 $\pm$    14.2 \\ 
2301163 & 14.3 & 2 &  & 141.03 &  20.82 &      7.8 &    160.5 $\pm$    20.5 \\ 
2302092 & 14.4 & 123 &  & 146.33 &  52.18 &      7.5 &   2065.1 $\pm$   276.6 \\ 
2302548 & 13.6 & 123 & 988 & 133.06 &  10.38 &     21.9 &    693.7 $\pm$    31.7 \\ 
2302889 & 13.6 & 123 &  & 144.26 &  26.77 &      7.5 &    174.8 $\pm$    23.5 \\ 
2304320 & 13.8 & 123 &  & 133.93 &  16.54 &     10.8 &    296.3 $\pm$    27.5 \\ 
2304655 & 12.4 & 123 &  & 132.02 &   0.54 &      7.5 &     26.1 $\pm$     3.5 \\ 
2304850 & 12.3 & 1 &  & 133.29 &   2.19 &      7.3 &     77.8 $\pm$    10.7 \\ 
2305372 & 13.8 & 123 &  & 131.52 &   1.40 &    522.3 &  59221.4 $\pm$   113.4 \\ 
2305543 & 12.5 & 23 &  & 131.62 &   0.68 &    266.6 &  23763.1 $\pm$    89.1 \\ 
2306740 & 13.5 & 123 &  & 138.72 &  10.31 &   2513.8 & 137639.9 $\pm$    54.8 \\ 
2307199 & 14.0 & 123 & 151 & 132.83 &  13.45 &     30.0 &   1171.6 $\pm$    39.1 \\ 
2307415 & 13.0 & 123 &  & 133.38 &  13.12 &      9.3 &    175.9 $\pm$    18.8 \\ 
2307533 & 12.6 & 123 &  & 131.57 &   0.54 &      7.2 &     27.7 $\pm$     3.8 \\ 
2308957 & 14.5 & 123 &  & 132.16 &   2.22 &    172.8 &  48760.9 $\pm$   282.2 \\ 
2309235 & 12.8 & 123 &  & 166.32 &  35.86 &      7.3 &    604.4 $\pm$    83.0 \\ 
2309318 & 15.2 & 123 &  & 178.96 &  66.11 &      7.3 &    857.2 $\pm$   117.7 \\ 
2309550 &  9.2 & 123 &  & 200.96 &  70.87 &      8.6 &    534.8 $\pm$    62.2 \\ 
2309587 & 13.9 & 123 &  & 132.10 &   1.84 &    512.9 &  57485.1 $\pm$   112.1 \\ 
2309719 & 12.9 & 123 & 1020 & 164.06 &  54.36 &    161.2 &   6708.6 $\pm$    41.6 \\ 
2423932 & 13.0 & 1 &  & 135.10 &   4.88 &      7.4 &    199.6 $\pm$    26.8 \\ 
2436772 & 14.2 & 2 &  & 133.21 &   6.35 &      7.9 & 377593.9 $\pm$ 47917.7 \\ 
2437452 & 17.0 & 123 &  & 134.56 &   7.24 &    417.5 &  95037.9 $\pm$   227.7 \\ 
2437488 & 15.0 & 123 &  & 141.84 &  14.47 &     17.3 &   1574.5 $\pm$    90.9 \\ 
2438070 & 13.8 & 123 & FP & 132.02 &   2.44 &     41.3 &    692.0 $\pm$    16.8 \\ 
2438264 & 14.2 & 123 & 440 & 136.07 &   4.97 &     20.7 &    626.9 $\pm$    30.3 \\ 
2438502 & 16.2 & 23 & 1003 & 138.85 &   8.36 &    122.5 &  21114.8 $\pm$   172.4 \\ 
2438513 & 14.0 & 123 &  & 139.06 &  12.18 &      9.5 &    396.0 $\pm$    41.5 \\ 
2440757 & 15.1 & 12 & FP & 131.68 &   1.43 &     14.4 &    368.2 $\pm$    25.5 \\ 
\enddata
\tablecomments{Table \ref{t1} is published in its entirety in the 
electronic edition of this journal.  A portion is shown here
for guidance regarding its form and content.}
\end{deluxetable}


\end{document}